\documentclass{jpsj2}

\title{A non-Hermitian critical point and the correlation length of strongly correlated quantum systems}
\author{\textsc{Yuichi Nakamura}$^{1}$\thanks{E-mail address: yuichi@iis.u-tokyo.ac.jp} and \textsc{Naomichi Hatano}$^{2}$\thanks{E-mail address: hatano@iis.u-tokyo.ac.jp}}
\inst{$^{1}$Department of Physics, University of Tokyo, Komaba, Meguro, Tokyo, 153-8505\\
$^{2}$Institute of Industrial Science, University of Tokyo, Komaba, Meguro, Tokyo, 153-8505}
\abst{We study a non-Hermitian generalization of quantum systems in which an imaginary vector potential is added to the momentum operator. In the tight-binding approximation, we make the hopping energy asymmetric in the Hermitian Hamiltonian. In a previous article, we conjectured that the non-Hermitian critical point where the energy gap vanishes is equal to the inverse correlation length of the Hermitian system and we confirmed the conjecture for two exactly solvable systems. In this article, we present more evidence for the conjecture.
We also argue the basis of our conjecture by noting the dispersion relation of the elementary excitation.
}

\kword{Non-Hermitian quantum mechanics, correlation length, dispersion relation, isotropic \textit{XY} chain, Hubbard model, $S=1/2$ antiferromagnetic \textit{XXZ} chain, Majumdar-Ghosh model}

\begin{document}
\maketitle

\section{Introduction} 

In this paper, we study a non-Hermitian generalization of strongly correlated quantum systems in which an imaginary vector potential $i\vec{g}$ (where $\vec{g}$ is a real vector) is added to the momentum operator.
The non-Hermitian kinetic energy in the continuous space is given by~\cite{Hatano}
\begin{equation}
\mathcal{H}_{\text{k}}=\frac{(-i\hbar\vec{\nabla}+i\vec{g})^2}{2m}.
\end{equation}
Its second-quantized form within the tight-binding approximation is given by
\begin{equation}
\mathcal{H}_{\text{k}} = -t\sum_{\nu=1}^{d}\sum_{\vec{x}} \left(e^{g_{\nu}(\vec{x})}c_{\vec{x}+\vec{e}_{\nu}}^\dag c_{\vec{x}}+e^{-g_{\nu}(\vec{x})}c_{\vec{x}}^\dag c_{\vec{x}+\vec{e}_{\nu}}\right).\label{non-Herm-generalization_d-dimension}
\end{equation}
In this article, we focus on the one-dimensional case with a constant imaginary vector potential and hereafter use 
\begin{equation}
\mathcal{H}_{\text{k}} = -t\sum_{x} (e^{g}c_{x+1}^\dag c_{x}+e^{-g}c_{x}^\dag c_{x+1}),\label{non-Herm-generalization}
\end{equation}
where $g$ is a real constant; in other words, we make the hopping energy asymmetric. More generally, we multiply the right hopping energy $c_{x+n}^{\dag}c_{x}$ by $e^{ng}$ and the left hopping energy $c_{x}^{\dag}c_{x+n}$ by $e^{-ng}$ in the original Hermitian Hamiltonian.  

The purpose of the non-Hermitian generalization is to obtain a \textit{length} scale inherent in the wave function of the Hermitian system only from the non-Hermitian energy spectrum.
The non-Hermitian generalization was first applied to the one-electron Anderson model by Hatano and Nelson~\cite{Hatano}. Their model is, in one dimension,
\begin{align}
\mathcal{H}_{\text{random}}(g)=&t\sum_{x=1}^{L}\left(e^g| x+1\rangle\langle x|+e^{-g}| x\rangle\langle x+1|\right)+\sum_{x=1}^{L}V_{x}|x\rangle\langle x|,
\label{Anderson}
\end{align}
where $V_{x}$ is a random potential at site $x$ and we require the periodic boundary condition.
As we increase the non-Hermiticity $g$, a pair of eigenvalues collide at a point $g=g_{\mathrm{c}}$ and then become complex. It was revealed~\cite{Hatano} that the non-Hermitian critical point $g_{\mathrm{c}}$ is equal to the inverse localization length of the eigenfunction of the original Hermitian Hamiltonian. 

We here apply the non-Hermitian generalization to systems without randomness but with interactions. We conjectured in the previous article~\cite{Nakamura} that we can obtain the \textit{correlation} length from the non-Hermitian generalization of strongly correlated quantum systems; the non-Hermitian critical point $g=g_{\text{c}}$ where the energy gap from the ground state collapses is equal to the inverse correlation length of the ground state of the Hermitian system. We confirmed the conjecture for two one-dimensional exactly solvable systems: the Hubbard model in the half-filled case and the $S=1/2$ antiferromagnetic \textit{XXZ} chain in the Ising-like region.

In the present article, we give further evidence for the conjecture in various levels of certainty. We also clarify the reason why we can obtain the inverse correlation length by our non-Hermitian generalization, by noting the dispersion relation of the elementary excitation. We show that the non-Hermitian generalization is actually equivalent to replacing $k$ with $k+ig$ in the dispersion relation of the elementary excitation; we thus seek a zero of the dispersion relation by computing the non-Hermitian critical point $g_{\text{c}}$. 

In \S~2, we confirm our conjecture for the following exactly solvable systems in one dimension: the $S=1/2$ ferromagnetic isotropic \textit{XY} chain in a magnetic field in \S~2.1; the half-filled Hubbard model in \S~2.2; the $S=1/2$ antiferromagnetic \textit{XXZ} chain in the Ising-like region~in \S~2.3; the Majumdar-Ghosh model in \S~2.4.

In \S~3, we numerically analyze non-Hermitian models of finite size $L$. We calculate the non-Hermitian ``critical" point $g_{\text{c}}(L)$ where the energy of the eigenstate corresponding to the ground state in the limit $L\to\infty$ becomes complex; we then obtain an extrapolated estimate $g_{\text{c}}(\infty)$. We numerically confirm that the estimate $g_{\text{c}}(\infty)$ and the inverse correlation length of the Hermitian systems are consistent for the Hubbard model and for the \textit{XXZ} model. We also analyze an unsolved model, namely $S=1/2$ antiferromagnetic Heisenberg chain with nearest- and next-nearest-neighbor interactions.

In the summary, we conclude our discussions by giving a remark on applying our non-Hermitian generalization.

\section{Non-Hermitian analysis of solvable models}

For four exactly solvable one-dimensional systems, we here confirm our conjecture that the non-Hermitian critical point is equal to the inverse correlation length of the Hermitian system.
We also reveal in this section that the non-Hermitian generalization results in replacing the real momentum $k$ with a complex one $k+ig$ in the dispersion relation of the elementary excitation. That is, the non-Hermitian generalization makes the Hermitian Hamiltonian of the form
\begin{equation}
\mathcal{H}=\sum_{-\pi<k<\pi}\epsilon(k)\eta_{k}^{\dag}\eta_{k} \label{Hermitian_Hamiltonian}
\end{equation}
transformed to
\begin{equation}
\mathcal{H}(g)=\sum_{-\pi<k<\pi}\epsilon(k+ig)\eta_{k}^{\dag}\eta_{k}.
\label{dispersion_non-Hermitian}
\end{equation}
Finding the non-Hermitian critical point where the energy gap above the ground state collapses is thereby equivalent to seeking a zero of the dispersion relation in the complex momentum plane.
The zero of the dispersion relation may be related to the inverse correlation length of the Hermitian system; see Appendix~B.
We confirm the equivalence for the isotropic \textit{XY} chain in \S~2.1 and for systems solved by the Bethe-ansatz method in \S~2.2 and \S~2.3.
We thereby show for the three models that the non-Hermitian critical point $g_{\text{c}}$ where the energy gap vanishes is rigorously equal to the inverse correlation length.

We also analyze in \S~2.4 the Majumdar-Ghosh model, for which we do not know the energy gap exactly; only approximate estimates are known. 
We still show that the non-Hermitian critical point $g_{\text{c}}$ where the approximate estimates of the energy gap vanish are equal to the inverse correlation length calculated by finite-size scaling of the correlation function of the ground state of the Hermitian Majumdar-Ghosh model.
We also argue that the non-Hermitian critical point $g_{\text{c}}$ obtained from the approximate estimates is the exact one.

\subsection{The isotropic \textit{XY} chain in a magnetic field}
We first consider the non-Hermitian generalization of the $S=1/2$ ferromagnetic isotropic \textit{XY} chain in a magnetic field. The Hermitian Hamiltonian of this model is 
\begin{equation}
\mathcal{H}_{XY}=-J\sum_{l=1}^{L}\left(S_{l}^{x}S_{l+1}^{x}+S_{l}^{y}S_{l+1}^{y}\right)-h\sum_{l=1}^{L}S_{l}^{z},
\label{XX_Hamiltonian}
\end{equation}
where we set $J>0$.
The Hamiltonian~(\ref{XX_Hamiltonian}) is transformed into
\begin{equation}
\mathcal{H}_{XY}=-\frac{J}{2}\sum_{l=1}^{L}(c_{l+1}^{\dag}c_{l}+c_{l}^{\dag}c_{l+1})+\frac{hL}{2}-h\sum_{l=1}^{L}c_{l}^{\dag}c_{l}
\label{XX_space_Hamiltonian}
\end{equation}
by the Jordan-Wigner transformation
\begin{equation}
c_{j}=(-2)^{j-1}S_{1}^{z}S_{2}^{z}\dots S_{j-1}^{z}S_{j}^{-},\hspace{9pt}c_{j}^{\dag}=(-2)^{j-1}S_{1}^{z}S_{2}^{z}\dots S_{j-1}^{z}S_{j}^{+}.\label{Jordan-Wigner}
\end{equation}
Applying the non-Hermitian generalization of the form~(\ref{non-Herm-generalization}) to the Hamiltonian~(\ref{XX_space_Hamiltonian}), we have 
\begin{equation}
\mathcal{H}_{XY}(g)=-\frac{J}{2}\sum_{l=1}^{L}(e^{g}c_{l+1}^{\dag}c_{l}+e^{-g}c_{l}^{\dag}c_{l+1})+\frac{hL}{2}-h\sum_{l=1}^{L}c_{l}^{\dag}c_{l}.
\label{XX_space_non-Herm_Hamiltonian}
\end{equation}
By the inverse Jordan-Wigner transformation, the Hamiltonian~(\ref{XX_space_non-Herm_Hamiltonian}) is transformed back into 
\begin{align}
\mathcal{H}_{XY}(g)=&-\frac{J}{2}\sum_{l=1}^{L}\left[e^{g}S_{l}^{-}S_{l+1}^{+}+e^{-g}S_{l}^{+}S_{l+1}^{-}\right]-h\sum_{l=1}^{L}S_{l}^{z}\notag\\
=&-J\sum_{l=1}^{L}\left[\cosh g\left(S_{l}^{x}S_{l+1}^{x}+S_{l}^{y}S_{l+1}^{y}\right)-i\sinh g\left(S_{l}^{y}S_{l+1}^{x}+S_{l}^{x}S_{l+1}^{y}\right)\right]-h\sum_{l=1}^{L}S_{l}^{z}.
\end{align}

We can immediately diagonalize the non-Hermitian Hamiltonian~(\ref{XX_space_non-Herm_Hamiltonian}) in the momentum space in the form
\begin{equation}
\mathcal{H}_{XY}(g)=\sum_{-\pi<k<\pi}\epsilon(k+ig)c_{k}^{\dag}c_{k}+\frac{hL}{2}
\label{XX_momentum_Hamiltonian}
\end{equation}
with the Fourier transformation
\begin{equation}
c_{k}=\frac{1}{\sqrt{L}}\sum_{l=1}^{L}e^{-ikl}c_{l},\hspace{9pt}c_{k}^{\dag}=\frac{1}{\sqrt{L}}\sum_{l=1}^{L}e^{ikl}c_{l}^{\dag},
\label{Fourier}
\end{equation}
where $\epsilon(k)\equiv-J\cos k-h$.
The non-Hermitian generalization thus shifts the momentum $k$ by $ig$ in the dispersion relation of the elementary excitation.

We define the non-Hermitian critical point $g_{\text{c}}$ as the point where the energy gap above the ground state vanishes. 
For $h>J$, all eigenvalues are real at the Hermitian point $g=0$ and $\epsilon(0)$ gives a finite energy gap~(Fig~1.~(a)). As we turn on the non-Hermiticity $g$, all eigenvalues except for  $k=0$ and  $\pm\pi$ immediately spread into the complex $k$ plane~(Fig. 1~(b)).
\begin{figure}
 \begin{center}
  (a)\includegraphics[width=7cm,clip]{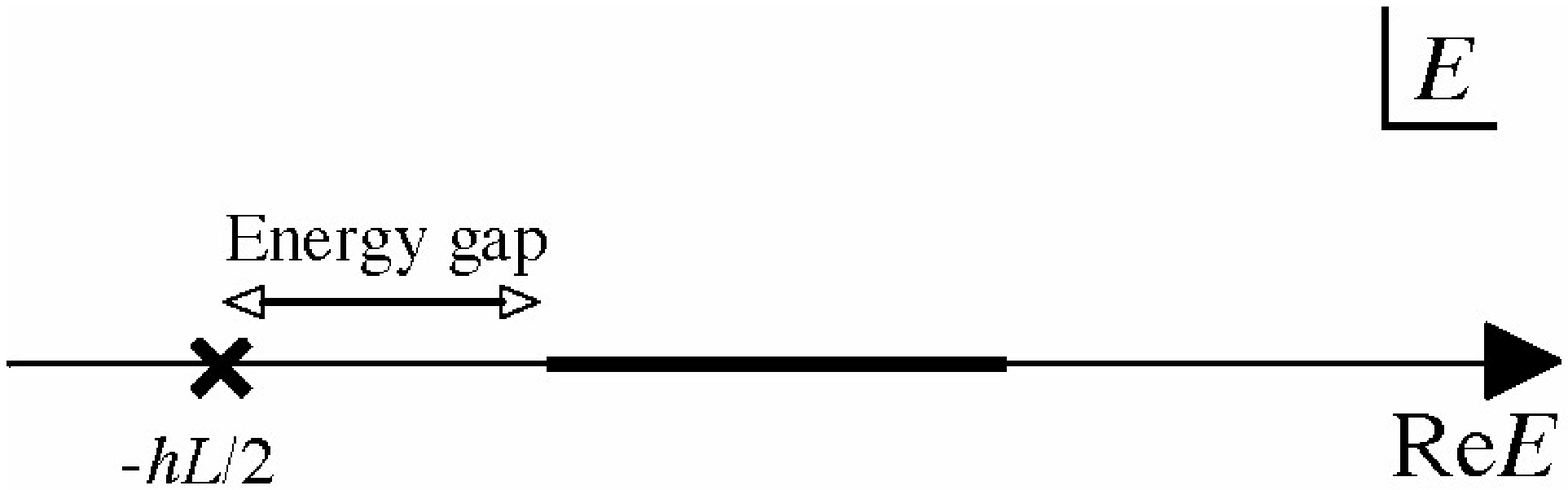}
  (b)\includegraphics[width=7cm,clip]{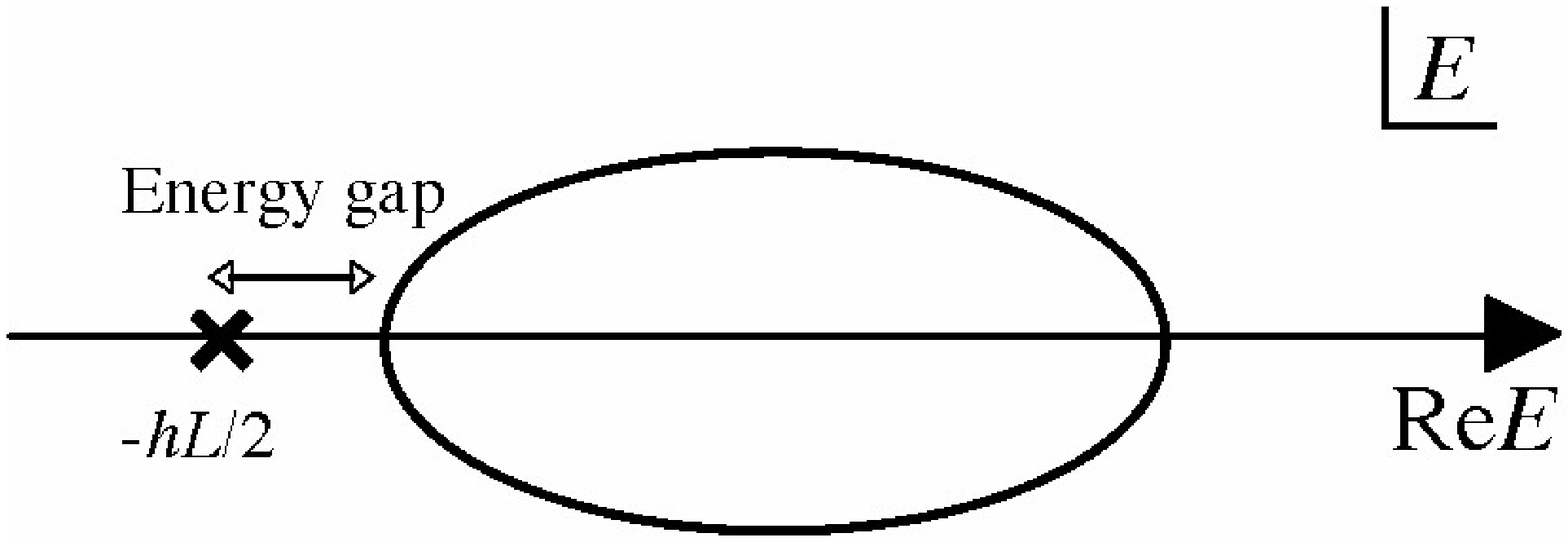}\label{XX_eigen_disti}
  \caption[]{The eigenvalue distributions of the non-Hermitian isotropic \textit{XY} chain for $h>J$ for (a) $g=0$ and (b) $0<g<g_{\text{c}}$. The symbol \textbf{$\times$} indicates the ground state.}
 \end{center}
 \end{figure}
 The energy gap $\Delta E(g)$ in the limit $L\to\infty$ is given by the elementary excitation energy in the form
 \begin{equation}
\Delta E(g)=\epsilon(\pm\pi+ig)=h-J\cosh g.
\end{equation}
The energy gap vanishes at
\begin{equation}
g_{\text{c}}=\ln\left[\frac{h}{J}+\sqrt{\left(\frac{h}{J}\right)^2-1}\right],\label{XY_gc}
\end{equation}
which is the non-Hermitian critical point of the model. We stress here that eq.~(\ref{XY_gc}) gives a zero of the dispersion relation:~$\epsilon(\pm\pi+ig_{\text{c}})=0$. It is plausible that the zero of the dispersion relation in the complex $k$ plane gives a typical length scale of the system. Indeed, we have
\begin{equation}
\epsilon(\pm\pi+i/\xi)=0.\label{dispersion_XY}
\end{equation}

\begin{figure}
\begin{center}
  \includegraphics[width=7cm,clip]{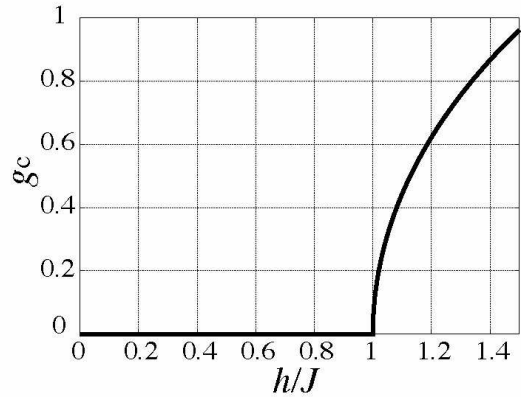} 
  \caption[]
  {The non-Hermitian critical point $g_{\text{c}}$ of the isotropic \textit{XY} chain. The ground-state critical point of the Hermitian system is $h/J=1$.}
  \label{gc-XX}
 \end{center}
 \end{figure}

Figure~\ref{gc-XX} shows the non-Hermitian critical point $g_{\text{c}}$ as a function of $h/J$. Hermitian system is gapless (the \textit{XY} phase) for $h<J$ and hence we have the non-Hermitian critical point $g_{\text{c}}=0$ in the region.
We can immediately confirm that the analytical expression of the non-Hermitian critical point $g_{\text{c}}$ is equal to the inverse correlation length of the Hermitian system, which was obtained by the quantum transfer matrix method~\cite{McCoy}. 

\subsection{The half-filled Hubbard chain}

We next consider the non-Hermitian generalization of the one-dimensional Hubbard model in the half-filled case:
\begin{equation}
\mathcal{H}_{\text{charge}}(g)=-t\sum_{l=1}^{L}\sum_{\sigma=\uparrow,\downarrow}\left(e^{g}c_{l+1,\sigma}^\dag c_{l,\sigma}+e^{-g}c_{l,\sigma}^\dag c_{l+1,\sigma}\right)+U\sum_{l=1}^{L}c_{l,\uparrow}^\dag c_{l,\uparrow} c_{l,\downarrow}^\dag c_{l,\downarrow},
\label{charge1}
\end{equation}
where $U$ is the on-site Coulomb repulsive interaction.
Fukui and Kawakami~\cite{Fukui} solved the non-Hermitian model~(\ref{charge1}) by the Bethe-ansatz method. They showed that we eliminate the charge gap, namely the Hubbard gap, as we increase the non-Hermiticity $g$. The fact is understood intuitively by computing the first-order perturbation of the Hamiltonian~(\ref{charge1}) with respect to $g$:
\begin{align}
&\mathcal{H}_{\text{charge}}(g)-\mathcal{H}_{\text{charge}}(0)\notag\\
&\cong -gt\sum_{l=1}^{L}(c_{l+1,\uparrow}^\dag c_{l,\uparrow}-c_{l,\uparrow}^\dag c_{l+1,\uparrow}+c_{l+1,\downarrow}^\dag c_{l,\downarrow}-c_{l,\downarrow}^\dag c_{l+1,\downarrow})=-ig(J_{\uparrow}+J_{\downarrow}),
 \label{charge2}
\end{align}
where $J_{\sigma}(\sigma=\uparrow,\downarrow)$ is the paramagnetic current operator defined by
\begin{equation}
J_{\sigma}\equiv-it\sum_{l=1}^{L}(c_{l+1,\sigma}^\dag c_{l,\sigma}-c_{l,\sigma}^\dag c_{l+1,\sigma}).\label{flow_spin}
\end{equation}
Hence $J_{\uparrow}+J_{\downarrow}$ is the charge current operator; see eq.~(\ref{spinon2}) below for comparison.
At the Hermitian point $g=0$ of the Hamiltonian~(\ref{charge1}) in the limit $L\to\infty$, the ground state is the Mott insulator for any finite $U$. As we increase the non-Hermiticity $g$, that is, as we try to produce the charge flow~(\ref{flow_spin}) in one direction, we expect that the ground state changes from the Mott insulator to an extended state, or a ``metallic" state at a point $g=g_{\text{c}}$; this is the non-Hermitian critical point where the Hubbard gap vanishes.

In order to solve the non-Hermitian Hamiltonian~(\ref{charge1}), we make the ansatz for the right eigenfunction $\Psi^{\text{R}}$, which is the same as in the Hermitian case:
\begin{equation}
\Psi^{\text{R}}(x_1,x_2,\dots,x_M | x_{M+1},\dots,x_N)=\sum_{P}[Q,P]\exp(i\sum_{j=1}^{N}k_{P_{j}}x_{Q_{j}}),
\label{Hubbard_Bethe_ansatz_wf_Nakamura}
\end{equation}
where $M$ and $N$ are the number of the down spins and the number of the electrons, respectively. 
The Bethe-ansatz wave function~(\ref{Hubbard_Bethe_ansatz_wf_Nakamura}) is different from that of Fukui and Kawakami's; their Bethe-ansatz wave function is given by replacing $k_j$ with $k_j-ig$ in eq.~(\ref{Hubbard_Bethe_ansatz_wf_Nakamura}). 
The difference lies only in the definition of $k_{j}$. We here use the form~(\ref{Hubbard_Bethe_ansatz_wf_Nakamura}) in order to compare the dispersion relation~(\ref{dispersion_non-Herm_Hubbard}) below with the Hermitian one. 
We put the down spins at $x_1,x_2,\dots,x_M$ and the up spins at $x_{M+1},\dots,x_N$. The symbols $P=(P_{1},P_{2},\dots,P_{N})$ and $Q=(Q_{1},Q_{2},\dots,Q_{N})$ are two permutations of the set $(1,2,\dots,N)$ with $1\leq x_{Q_{1}}\leq x_{Q_{2}}\leq \dots \leq x_{Q_{N}} \leq L$. The symbol $[Q,P]$ is a set of $N!\times N!$ coefficients depending on the two permutations $P$ and $Q$.
The ``dispersion relation" with respect to the quasimomentum $k_j$ is given by
\begin{equation}
\epsilon(k_j)=-2t\cos(k_j+ig),\label{dispersion_non-Herm_Hubbard}
\end{equation} 
where $k_j$ is a solution of the Bethe-ansatz equations:
\begin{align}
&\exp(iLk_{j})=\prod_{\beta=1}^{M}\frac{\sin(k_{j}+ig)-\Lambda_{\beta}+iU/4t}{\sin(k_{j}+ig)-\Lambda_{\beta}-iU/4t}\hspace{9pt}(j=1,\dots,N), \notag\\
&\prod_{j=1}^{N}\frac{\sin(k_{j}+ig)-\Lambda_{\alpha}+iU/4t}{\sin(k_{j}+ig)-\Lambda_{\alpha}-iU/4t}=
-\prod_{\beta=1}^{M}\frac{\Lambda_{\alpha}-\Lambda_{\beta}-iU/4t}{\Lambda_{\alpha}-\Lambda_{\beta}+iU/4t}\hspace{9pt}(\alpha=1,\dots,M).\label{non-Herm_Bethe1}
\end{align}
The non-Hermitian generalization thus results in replacing $k_j$ with $k_j+ig$ in the dispersion relation.

Fukui and Kawakami~\cite{Fukui} obtained an analytical expression of the non-Hermitian critical point $g_{\text{c}}$ in the limit $L\to\infty$, in the form
\begin{equation}
g_{\text{c}}=\text{arcsinh}(U/4t)+2i\int_{-\infty}^{\infty}\arctan\frac{\lambda+iU/4t}{U/4t}\sigma(\lambda)d\lambda,
\label{gc_Hubbard}
\end{equation}
where the distribution function $\sigma(\lambda)$ of the spin rapidity $\lambda$ is given by
\begin{equation}
\sigma(\lambda)=\frac{1}{2\pi}\int_{0}^{\infty}\text{sech}\left(\frac{U}{4t}\omega \right)\cos(\lambda\omega)J_0(\omega)d\omega
\end{equation}
with $J_0(\omega)$ the Bessel function of the first kind.
After some algebra in Appendix~A, we can show that the non-Hermitian critical point $g_{\mathrm{c}}$ is actually equal to the inverse correlation length $1/\xi$ due to the charge excitation of the Hermitian Hubbard model in the form
\begin{equation}
\frac{1}{\xi}=\text{arcsinh}\left(\frac{U}{4t}\right)-2\int_{0}^{\infty}\frac{J_{0}(\omega)\sinh((U/4t)\omega)}{\omega(1+e^{2(U/4t)\omega})}d\omega,
\label{xi_Hubbard}
\end{equation}
which was obtained by Stafford and Millis~\cite{Stafford} from finite-size scaling of the Drude weight.

We note here that eq.~(\ref{xi_Hubbard}) is equal to the imaginary part of the momentum at a zero of the dispersion relation of the charge excitation in the complex momentum space.
This is the same situation as in~(\ref{dispersion_XY}) for the \textit{XY} chain; see Appendix~B.1 for details. 

Figure~\ref{gap_g_diagram} exemplifies how the Hubbard gap collapses as we increase the non-Hermiticity $g$. The ground-state energy does not change and the Hubbard gap gradually decreases before it vanishes at $g=g_{\text{c}}$.
The way the energy gap collapses is different from that for the Anderson model discussed by Hatano and Nelson~\cite{Hatano}, where the energy gap decreases almost suddenly as $g$ gets close to $g=g_{\text{c}}$.
Unfortunately, it is difficult to know the ground-state properties of the non-Hermitian Hubbard model (\ref{charge1}) for $g>g_{\mathrm{c}}$. However, we expect that the ground-state energy becomes complex in the region $g>g_{\mathrm{c}}$ on the basis of finite-size data shown in \S~3.1.

\begin{figure}
\begin{center}
  \includegraphics[width=7cm,clip]{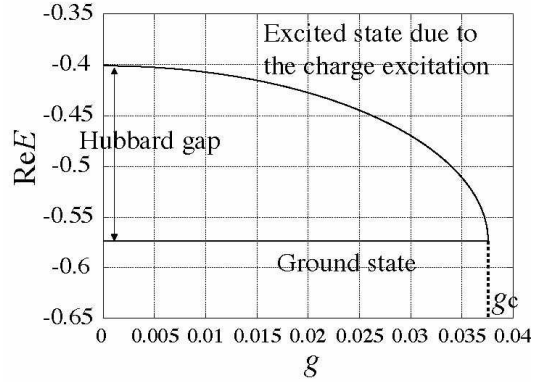} 
  \caption[]
  {The $g$ dependence of the Hubbard gap $\Delta$ for $U/t=2$. The non-Hermitian critical point is $g_{\text{c}}\cong 0.038.$}
  \label{gap_g_diagram}
 \end{center}
 \end{figure}

\subsection{The $S=1/2$ antiferromagnetic \textit{XXZ} chain}

We next develop a parallel discussion for a spin system, namely the $S=1/2$ antiferromagnetic \textit{XXZ} chain. The non-Hermitian Hamiltonian of the \textit{XXZ} chain is given by~\cite{Albertini}
\begin{equation}
\mathcal{H}_{XXZ}(g)=J\sum_{l=1}^{L}\left[\frac{1}{2}\left(e^{2g} S_{l}^{-}S_{l+1}^{+}+e^{-2g}S_{l}^{+}S_{l+1}^{-}\right)+\Delta S_{l}^{z}S_{l+1}^{z} \right]\label{XXZ-non-Hermitian}
\end{equation}
for $J>0$. We set $S_{\mathrm{tot}}^{z}=0$ hereafter.
The non-Hermitian Hamiltonian $\mathcal{H}_{XXZ}(g)$ in the case of $\Delta=1$ is derived as an effective Hamiltonian of the following non-Hermitian Hubbard model: 
\begin{equation}
\mathcal{H}_{\text{spin}}(g)=-t\sum_{l=1}^{L}(e^{g}c_{l+1,\uparrow}^\dag c_{l,\uparrow}+e^{-g}c_{l,\uparrow}^\dag c_{l+1,\uparrow}+e^{-g}c_{l+1,\downarrow}^\dag c_{l,\downarrow}+e^{g}c_{l,\downarrow}^\dag c_{l+1,\downarrow})+U\sum_{l=1}^{L}c_{l,\uparrow}^\dag c_{l,\uparrow} c_{l,\downarrow}^\dag c_{l,\downarrow}.\label{spinon1}
\end{equation}
We have the Hamiltonian~(\ref{XXZ-non-Hermitian}) with $\Delta=1$ by considering the second-order perturbation with respect to the kinetic energy around the degenerate ground state. 
Note the difference between eqs.~(\ref{charge1}) and (\ref{spinon1}); the first-order perturbation with respect to the non-Hermiticity $g$ gives 
\begin{equation}
\mathcal{H}_{\text{spin}}(g)-\mathcal{H}_{\text{spin}}(0)=-ig(J_{\uparrow}-J_{\downarrow}),
 \label{spinon2}
\end{equation}
where $J_{\uparrow}-J_{\downarrow}$ is the spin current operator.
Thus we expect that the non-Hermiticity $g$ induces a spin current and eliminates the spin gap.

Albertini \textit{et al.}\cite{Albertini} solved eq.~(\ref{XXZ-non-Hermitian}) in the massive region $\Delta>1$ in the subspace $S_{\mathrm{tot}}^{z}=0$. 
In order to solve the non-Hermitian Hamiltonian~(\ref{XXZ-non-Hermitian}), we make the following ansatz for the right eigenfunction $\Psi^{\text{R}}$, which is the same as in the Hermitian case:
\begin{equation}
\Psi^{\text{R}}(x_1,x_2,\dots,x_M)=\sum_{P}[Q,P]\exp(i\sum_{j=1}^{M}k_{P_{j}}x_{Q_{j}}),\label{XXZ_Bethe_wf_Nakamura}
\end{equation} 
where we put the down spins at $x_1,x_2,\dots,x_M$ $(1\leq x_1<x_2<\dots<x_M\leq L)$.
The Bethe-ansatz wave function~(\ref{XXZ_Bethe_wf_Nakamura}) is different from that of Albertini \textit{et al.}'s; their Bethe-ansatz wave function is given by replacing $k$ with $k-2ig$ in eq.~(\ref{XXZ_Bethe_wf_Nakamura}); see a comment below eq.~(\ref{Hubbard_Bethe_ansatz_wf_Nakamura}).

The ``dispersion relation" with respect to the quasimomentum $k_j$ is given by 
\begin{equation}
\epsilon(k_j)=J\left[\Delta+\cos(k_j+2ig)\right],\label{dispersion_non-Herm_XXZ}
\end{equation}
where $k_j$ is a solution of the Bethe-ansatz equations:
\begin{equation}
\exp(ik_{j}L)=(-1)^{M-1}\prod_{l(\neq j)}^{M}\frac{\exp[i(k_j+k_l)-4g]+1-2\Delta\exp(ik_j-2g)}{\exp[i(k_j+k_l)-4g]+1-2\Delta\exp(ik_l-2g)}\hspace{9pt}(j=1, 2,\cdots, M).\label{XXZ_eq0}
\end{equation}
The non-Hermitian generalization results in replacing $k_j$ by $k_j+2ig$ in the dispersion relation. 

Albertini \textit{et al.} obtained an analytical expression in the limit $L\to\infty$ of the non-Hermitian critical point $g_{\mathrm{c}}$ at which the spin gap vanishes, in the form~\cite{Albertini}
\begin{equation}
g_{\text{c}}=\frac{\gamma}{2}+\sum_{n=1}^{\infty}\frac{(-1)^n\tanh(n\gamma)}{n},\label{gc_XXZ}
\end{equation}
where $\gamma=\text{arccosh}\Delta$. Although Albertini \textit{et al.} did not point out the fact, the expression~(\ref{gc_XXZ}) is actually well known as the inverse correlation length~\cite{Baxter} at the Hermitian point $g=0$ obtained by the quantum transfer matrix method.
We also note that eq.~(\ref{gc_XXZ}) is related to the imaginary part of the momentum at a zero of the dispersion relation of the spinon excitation in the complex momentum space; see Appendix~B.2 for details.

It is again difficult to know the ground-state properties of the non-Hermitian $XXZ$ chain (\ref{XXZ-non-Hermitian}) for $g>g_{\mathrm{c}}$. However, we expect that the ground-state energy becomes complex in the region $g>g_{\mathrm{c}}$ on the basis of finite-size data shown in \S~3.2.

\subsection{The Majumdar-Ghosh model}

We now consider the non-Hermitian generalization of the Majumdar-Ghosh model~\cite{MG1, MG2} 
 \begin{equation}
\mathcal{H}_{\text{MG}}=J\sum_{l=1}^{L}[\frac{1}{2}(S_{l+1}^{+}S_{l}^{-}+S_{l}^{+}S_{l+1}^{-})+S_{l}^{z}S_{l+1}^{z}] +\frac{J}{2}\sum_{l=1}^{L}[\frac{1}{2}(S_{l+2}^{+}S_{l}^{-}+S_{l}^{+}S_{l+2}^{-})\notag+S_{l}^{z}S_{l+2}^{z}].
\label{MG_Hermitian_Hamiltonian}
\end{equation} 
The Hamiltonian~(\ref{MG_Hermitian_Hamiltonian}) has two-fold degenerate ground states and has a finite energy gap~\cite{Affleck}.
However, only variational estimates of the energy gap are known~\cite{Shastry, Caspers}.
We introduce the non-Hermitian Hamiltonian of the Majumdar-Ghosh model by referring to the non-Hermitian generalization of the \textit{XXZ} chain discussed in \S~2.3:
 \begin{align}
\mathcal{H}_{\text{MG}}(g)=&J\sum_{l=1}^{L}[\frac{1}{2}(e^{2g}S_{l+1}^{+}S_{l}^{-}+e^{-2g}S_{l}^{+}S_{l+1}^{-})+S_{l}^{z}S_{l+1}^{z}] \notag\\
+&\frac{J}{2}\sum_{l=1}^{L}[\frac{1}{2}(e^{4g}S_{l+2}^{+}S_{l}^{-}+e^{-4g}S_{l}^{+}S_{l+2}^{-})\notag+S_{l}^{z}S_{l+2}^{z}].\notag\\
\label{MG_Hamiltonian}
\end{align} 
The above Hamiltonian can be derived from the effective Hamiltonian of the non-Hermitian $t$-$t^{\prime}$-$U$ model as follows:
\begin{align}
\mathcal{H}_{t\text{-}t^{\prime}\text{-}U}(g)=&-t\sum_{l=1}^{L}(e^{g}c_{l+1,\uparrow}^\dag c_{l,\uparrow}+e^{-g}c_{l,\uparrow}^\dag c_{l+1,\uparrow}+e^{-g}c_{l+1,\downarrow}^\dag c_{l,\downarrow}+e^{g}c_{l,\downarrow}^\dag c_{l+1,\downarrow})\notag\\
&-t^{\prime}\sum_{l=1}^{L}(e^{2g}c_{l+2,\uparrow}^\dag c_{l,\uparrow}+e^{-2g}c_{l,\uparrow}^\dag c_{l+2,\uparrow}+e^{-2g}c_{l+2,\downarrow}^\dag c_{l,\downarrow}+e^{2g}c_{l,\downarrow}^\dag c_{l+2,\downarrow})\notag\\
&+U\sum_{l=1}^{L}c_{l,\uparrow}^\dag c_{l,\uparrow} c_{l,\downarrow}^\dag c_{l,\downarrow}.
\label{charge11}
\end{align}

Let us first calculate the correlation length of the Hermitian Majumdar-Ghosh model from finite-size scaling of the correlation function.
The two-fold degenerate ground states of the Hermitian Majumdar-Ghosh model~(\ref{MG_Hermitian_Hamiltonian}) are
\begin{align}
\Psi_{+}&=\frac{1}{\sqrt{2+4\cdot 2^{-L/2}}}(\Phi_{\text{I}}+\Phi_{\text{II}}),\notag\\
\Psi_{-}&=\frac{1}{\sqrt{2-4\cdot 2^{-L/2}}}(\Phi_{\text{I}}-\Phi_{\text{II}}),
\label{Hermitian_MG_wavefunc_1}
\end{align}
where wave functions $\Phi_{\text{I}}$ and $\Phi_{\text{II}}$ are defined by
\begin{align}
\Phi_{\text{I}}&=\phi_{1,2}\otimes\phi_{3,4}\otimes\dots\otimes\phi_{L-1,L},\notag\\
\Phi_{\text{II}}&=\phi_{2,3}\otimes\phi_{4,5}\otimes\dots\otimes\phi_{L,1}
\label{Hermitian_MG_wavefunc_2}
\end{align}
with $\phi_{i,j}$ denoting the singlet state of a pair of spins at sites $i$ and $j$:
\begin{equation*}
\phi_{i,j}=\frac{1}{\sqrt{2}}(|\uparrow_{i}\downarrow_{j}\rangle-|\downarrow_{i}\uparrow_{j}\rangle).
\end{equation*}
Note that $\Phi_{\text{I}}$ and $\Phi_{\text{I}}$ are not orthogonal: $\displaystyle\langle\Phi_{\text{I}}|\Phi_{\text{II}}\rangle=\langle\Phi_{\text{II}}|\Phi_{\text{I}}\rangle=2^{-L/2+1}$. The states~(\ref{Hermitian_MG_wavefunc_1}), on the other hand, are orthonormal. 

The correlation functions $\displaystyle\langle S^{z}_{0}S^{z}_{r}\rangle_{\pm}=\langle\Psi_{\pm} |S^{z}_{0}S^{z}_{r}|\Psi_{\pm}\rangle$ with respect to the two-fold degenerate ground states $\Psi_{+}$ and $\Psi_{-}$ of the system of size $L$ are given by
\begin{align}
\langle S^{z}_{0}S^{z}_{1}\rangle_{\pm}&=\dfrac{-1\mp 2^{-L/2+2}}{2\pm 2^{-L/2+2}}=-\frac{1}{2}\mp 2^{-L/2}+\text{O}\left((2^{-L/2})^2\right),\notag\\
\langle S^{z}_{0}S^{z}_{2i}\rangle_{\pm}&=\dfrac{\pm 2^{-L/2}}{1\pm 2^{-L/2+1}}=\pm 2^{-L/2}+\text{O}\left((2^{-L/2})^2\right)\hspace{9pt}(\text{for}~i\geq 1),\notag\\
\langle S^{z}_{0}S^{z}_{2i+1}\rangle_{\pm}&=\dfrac{\mp 2^{-L/2}}{1\pm 2^{-L/2+1}}=\mp 2^{-L/2}+\text{O}\left((2^{-L/2})^2\right)\hspace{9pt}(\text{for}~i\geq 1).
\label{Hermitian_MG_cor_express}
\end{align}
Assuming finite-size correction of the correlation function in the form 
\begin{equation}
\langle S^{z}_{0}S^{z}_{r}\rangle_{L}=\langle S^{z}_{0}S^{z}_{r}\rangle_{\infty}+\text{O}\left(\exp(-L/\xi)\right),
\end{equation}
we obtain the correlation length
\begin{equation}
\xi=\frac{2}{\ln 2}.
\end{equation}

We next calculate the non-Hermitian critical point $g_{\text{c}}$ where approximate estimates of the energy gap vanishes.
 
\begin{figure}
  \begin{center}
  \includegraphics[width=8cm,clip]{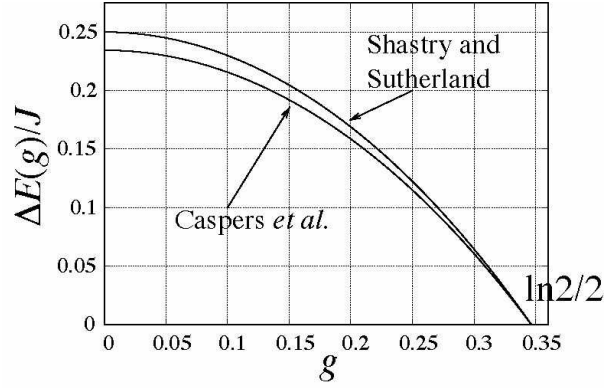} 
  \caption[]{The non-Hermiticity dependence of the approximate estimates of the energy gap calculated by Shastry and Sutherland's approach and by Caspers \textit{et al.}'s approach.}
\label{dispersion_MG}
 \end{center}
 \end{figure}

The dispersion relation was first obtained by Shastry and Sutherland~\cite{Shastry} with a trial wave function and then by Caspers \textit{et al.}~\cite{Caspers} with a variational wave function.
The dispersion relation of the Hermitian Majumdar-Ghosh model given by Shastry and Sutherland~\cite{Shastry} is
\begin{equation}
\epsilon_\text{S}(k)=J\left(\frac{5}{4}+\cos k\right);\label{dispersion_Shastry}
\end{equation}
the excitation energy at $k=\pm\pi$ determines the energy gap $\Delta E_{\text{S}}$. When we turn on the non-Hermiticity $g$ as in eq.~(\ref{MG_Hamiltonian}), the momentum $k$ in eq.~(\ref{dispersion_Shastry}) is replaced by $k+2ig$. 
The dependence of the energy gap $\Delta E_{\text{S}}(g)$ on the non-Hermiticity $g$ is 
\begin{equation}
\Delta E_{\text{S}}(g)=\epsilon_\text{S}(\pm\pi+2ig)=J\left(\frac{5}{4}-\cosh(2g)\right).\label{gap_Shastry}
\end{equation}
As shown in Fig.~\ref{dispersion_MG}, the non-Hermitian critical point $g_{\text{c}}$ where the energy gap $\Delta E_{\text{S}}$ vanishes is $g_{\text{c}}=\ln 2/2$.
 
The variational form of the dispersion relation given by Caspers \textit{et al.}~\cite{Caspers} is 
\begin{equation}
\epsilon_\text{C}(k)=\frac{(60r+34)\cos 2k+(139r+170)\cos k-(475r-731)}{(8r+4)\cos 2k-(28r+70)\cos k-(52r-86)}\label{dispersion_Caspers}
\end{equation}
with
\begin{equation}
r=\frac{\sqrt{2\cos 2k-20\cos k+43}}{5+4\cos k};
\end{equation}
the excitation energy at $k=\pm\pi$ determines the energy gap $\Delta E_{\text{C}}$.
When we turn on the non-Hermiticity $g$, the momentum $k$ in eq.~(\ref{dispersion_Caspers}) is again replaced by $k+2ig$.
The dependence of the energy gap $\Delta E_\text{C}(g)$ on the non-Hermiticity $g$ is 
\begin{equation}
\Delta E_{\text{C}}(g)=\frac{(60R+34)\cosh(4g)-(139R+170)\cosh(2g)-(475R-731)}{(8R+4)\cosh(4g)+(28R+40)\cosh(2g)-(52R-86)}\label{gap_Caspers}
\end{equation}
with
\begin{equation}
R=\sqrt{\frac{2\cosh(4g)+20\cosh(2g)+43}{(5-4\cosh(2g))^2}}.
\end{equation}
As shown in Fig.~\ref{dispersion_MG}, the non-Hermitian critical point $g_{\text{c}}$ is $g_{\text{c}}=\ln 2/2$ again. In both cases, the non-Hermitian critical point $g_{\text{c}}$ is equal to the inverse correlation length $\xi^{-1}=\ln 2/2$. 

We now give a piece of evidence that the  non-Hermitian critical point is exactly $g_{\text{c}}=\ln 2/2$ by showing that the ground-state property dramatically changes at $g=\ln 2/2$.
The non-Hermitian Majumdar-Ghosh model~(\ref{MG_Hamiltonian}) has two-fold degenerate dimer ground states as in the Hermitian case. 
The right eigenfunctions $|\Psi_{\pm}^{\text{R}}\rangle$ and the left eigenfunctions $\langle\Psi_{\pm}^{\text{L}}|$ of the ground states of the system of size $L$ are given by
\begin{align}
|\Psi_{\pm}^{\text{R}}\rangle&=\frac{1}{\sqrt{2\pm 2\dfrac{e^{gL}+e^{-gL}}{2^{L/2}}}}(|\Phi_{\text{I}}^{\text{R}}\rangle\pm|\Phi_{\text{II}}^{\text{R}}\rangle),\notag\\
\langle\Psi_{\pm}^{\text{L}}|&=\frac{1}{\sqrt{2\pm 2\dfrac{e^{gL}+e^{-gL}}{2^{L/2}}}}(\langle\Phi_{\text{I}}^{\text{L}}|\pm\langle\Phi_{\text{II}}^{\text{L}}|),\label{MG_wavefunc_1}
\end{align}
where the wave functions $|\Phi_{\text{I, II}}^{\text{R}}\rangle$ and $\langle\Phi_{\text{I, II}}^{\text{L}}|$ are defined by
\begin{align}
|\Phi_{\text{I}}^{\text{R}}\rangle&=|\phi_{1,2}^{\text{R}}\rangle\otimes|\phi_{3,4}^{\text{R}}\rangle\otimes\dots\otimes|\phi_{L-1,L}^{\text{R}}\rangle,\notag\\
|\Phi_{\text{II}}^{\text{R}}\rangle&=|\phi_{2,3}^{\text{R}}\rangle\otimes|\phi_{4,5}^{\text{R}}\rangle\otimes\dots\otimes|\phi_{L,1}^{\text{R}}\rangle,\notag\\
\langle\Phi_{\text{I}}^{\text{L}}|&=\langle\phi_{1,2}^{\text{L}}|\otimes\langle\phi_{3,4}^{\text{L}}|\otimes\dots\otimes\langle\phi_{L-1,L}^{\text{L}}|,\notag\\
\langle\Phi_{\text{II}}^{\text{L}}|&=\langle\phi_{2,3}^{\text{L}}|\otimes\langle\phi_{4,5}^{\text{L}}|\otimes\dots\otimes\langle\phi_{L,1}^{\text{L}}|\label{MG_wavefunc_2}
\end{align}
with $|\phi^{\text{R}}_{i,j}\rangle$ and $\langle\phi^{\text{L}}_{i,j}|$ denoting weighted singlet states of a pair of spins at sites $i$ and $j$:
\begin{align}
|\phi^{\text{R}}_{i,j}\rangle=\frac{1}{\sqrt{2}}(e^{-g}|\uparrow_{i}\downarrow_{j}\rangle-e^{g}|\downarrow_{i}\uparrow_{j}\rangle),\notag\\
\langle\phi^{\text{L}}_{i,j}|=\frac{1}{\sqrt{2}}(e^{g}\langle\uparrow_{i}\downarrow_{j}|-e^{-g}\langle\downarrow_{i}\uparrow_{j}|).\label{MG_wavefunc_3}
\end{align}
Equation~(\ref{MG_wavefunc_2}) is consistent with eq.~(\ref{Hermitian_MG_wavefunc_2}) transformed by a many-body version of the imaginary gauge transformation~\cite{Hatano}
\begin{equation}
\psi(x_1, x_2, \dots; g)=e^{g\sum_{i}x_i}\psi(x_1, x_2, \dots; 0).
\end{equation}
Note that $\langle\Phi_{\text{I}}^{\text{L}}|\Phi_{\text{I}}^{\text{R}}\rangle=\langle\Phi_{\text{II}}^{\text{L}}|\Phi_{\text{II}}^{\text{R}}\rangle=1$ and $\displaystyle\langle\Phi_{\text{I}}^{\text{L}}|\Phi_{\text{II}}^{\text{R}}\rangle=\langle\Phi_{\text{II}}^{\text{L}}|\Phi_{\text{I}}^{\text{R}}\rangle=\left(e^{gL}+e^{-gL}\right)/2^{L/2}$, but $|\Psi_{\pm}^{\text{R}}\rangle$ and $\langle\Psi_{\pm}^{\text{L}}|$ are bi-orthonormal.

\begin{figure}
  \begin{center}
(a)  \includegraphics[width=7cm,clip]{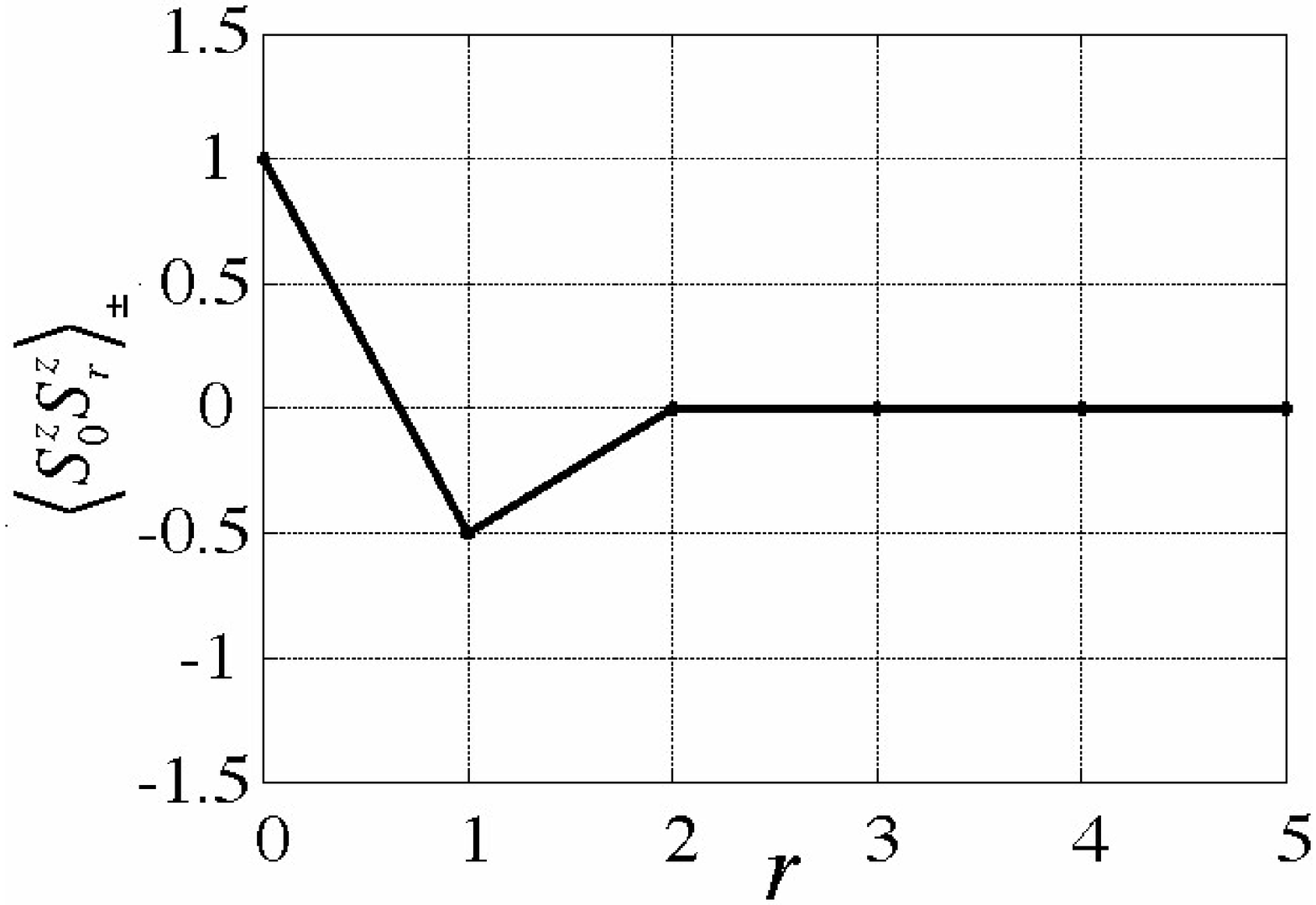} \\(b)\includegraphics[width=7cm,clip]{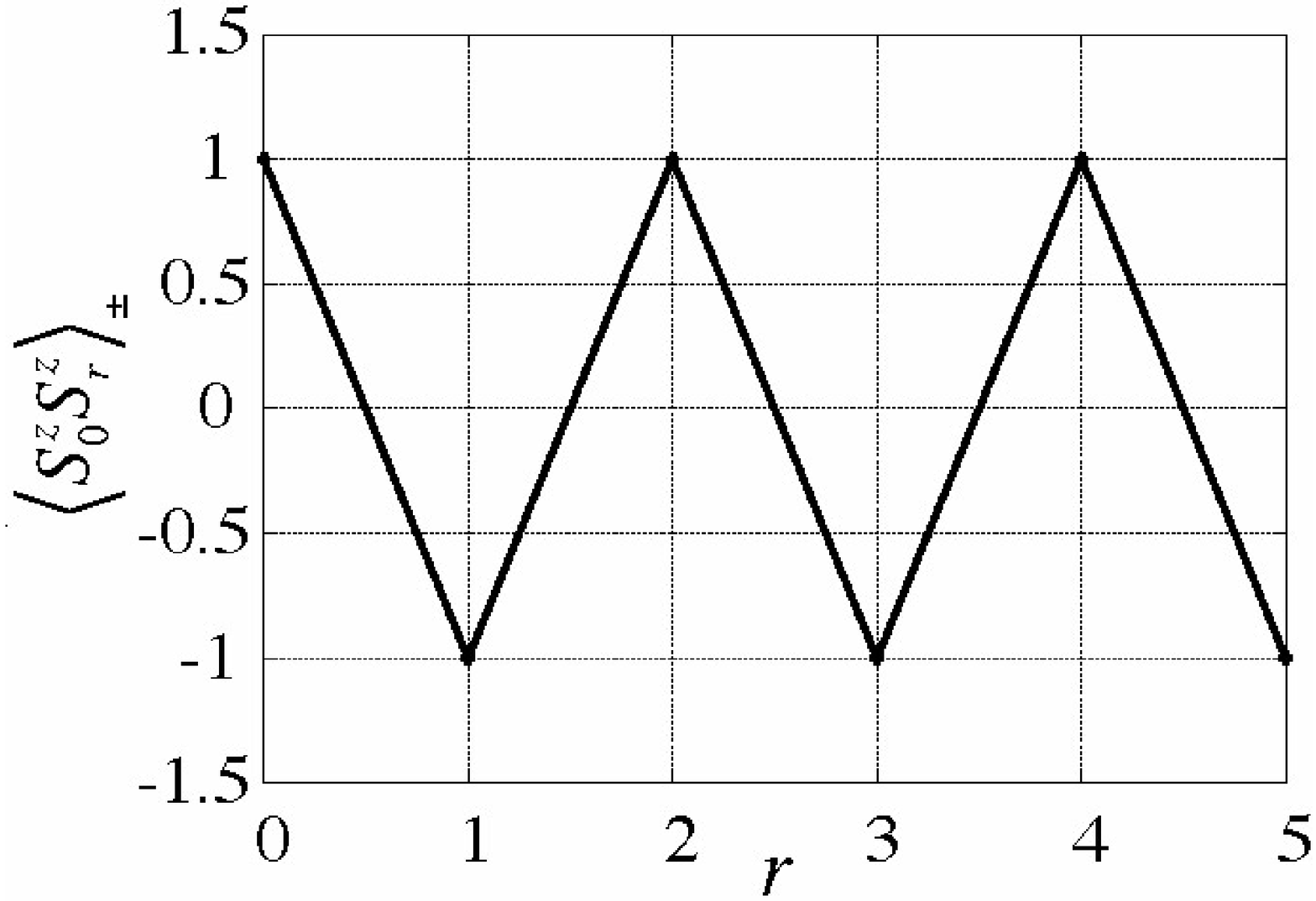} \\
  \caption[]{The correlation function $\langle S^{z}_{0}S^{z}_{r}\rangle_{\pm}$ of the non-Hermitian Majumdar-Ghosh model of infinite size in the regions (a) $g<\ln 2/2$ and (b) $g>\ln 2/2$.}
  \label{MG_cor}
 \end{center}
 \end{figure}

The correlation functions of the non-Hermitian system, 
$\langle S^{z}_{0}S^{z}_{r}\rangle_{\pm}=\langle\Psi^{\text{L}}_{\pm} |S^{z}_{0}S^{z}_{r}|\Psi^{\text{R}}_{\pm}\rangle$
with respect to the two-fold degenerate ground states $|\Psi_{+}^{\text{R}}\rangle$ and $|\Psi_{-}^{\text{R}}\rangle$ are obtained in the forms
\begin{align}
\langle S^{z}_{0}S^{z}_{1}\rangle_{\pm}&=\dfrac{-1\mp 2\dfrac{e^{gL}+e^{-gL}}{2^{L/2}}}{2\pm 2\dfrac{e^{gL}+e^{-gL}}{2^{L/2}} },\notag\\\notag\\
\langle S^{z}_{0}S^{z}_{2i}\rangle_{\pm}&=\dfrac{\pm\dfrac{e^{gL}+e^{-gL}}{2^{L/2}}}{2\pm 2\dfrac{e^{gL}+e^{-gL}}{2^{L/2}} }\hspace{9pt}(\text{for}~i\geq 1),\notag\\\notag\\
\langle S^{z}_{0}S^{z}_{2i+1}\rangle_{\pm}&=\dfrac{\mp\dfrac{e^{gL}+e^{-gL}}{2^{L/2}}}{2\pm 2\dfrac{e^{gL}+e^{-gL}}{2^{L/2}} }\hspace{9pt}(\text{for}~i\geq 1) . 
\label{MG_cor_express}
\end{align}
Figure~\ref{MG_cor} shows the correlation function in the limit $L\to\infty$ in the region $0<g<\ln 2/2$ and in the region $g>\ln 2/2$, respectively. The ground state is dimerized in the region $0<g<\ln 2/2$ and is an extended state in the region $g>\ln 2/2$. The phase transition from the dimer state to the extended state reminds us of the localization-delocalization transition of the non-Hermitian random Anderson model discussed by Hatano and Nelson~\cite{Hatano}. The phase transition point $g_{\text{c}}=\ln 2/2$ is then naturally regarded as the non-Hermitian critical point.
We thus confirm from the above discussions that the non-Hermitian critical point is equal to the inverse correlation length of the Hermitian systems for the Majumdar-Ghosh model.

\section{Numerical analysis of non-Hermitian models}

In the previous section, we discussed the non-Hermitian generalization of exactly solvable models and we confirmed the conjecture that the non-Hermitian critical point $g_{\text{c}}$ where the energy gap vanishes is equal to the inverse correlation length of the Hermitian system. We now numerically show that the inverse correlation length is consistent with the extrapolated estimate $g_{\text{c}}(\infty)$ of finite-size data $g_{\text{c}}(L)$ where an eigenvalue which corresponds to the ground state in the limit $L\to\infty$, becomes complex. We show the above for the Hubbard model in \S~3.1, for the $S=1/2$ antiferromagnetic \textit{XXZ} model in \S~3.2 and for a frustrated quantum spin chain in \S~3.3. Although we do not know the correlation length of the $S=1/2$ antiferromagnetic Heisenberg chain with nearest- and next-nearest-neighbor interactions, we show in \S~3.3 that the numerical estimate $g_{\text{c}}(\infty)$ is consistent with the ground-state phase diagram.

\subsection{The non-Hermitian Hubbard chain} 
We first analyze the non-Hermitian Hubbard models~(\ref{charge1}) and~(\ref{spinon1}) of size $L$.
We define the non-Hermitian ``critical" point $g_{\text{c}}(L)$ of a finite system as the point where the energy gap between the ground state and a low-lying excited state vanishes and beyond which the ground-state energy becomes complex. We here show that the extrapolated estimate $g_{\text{c}}(\infty)$ of finite-size data $g_{\text{c}}(L)$ is close to the exact value of the correlation length. 
 
We first use the non-Hermitian generalization of the form~(\ref{charge1}) in the subspace $S_{\text{tot}}^{z}=0$,  which eliminates the charge gap. All eigenvalues are real at the Hermitian point $g=0$. Upon increasing $g$, a pair of eigenvalues move on the real axis. They spread into the complex plane when $g$ exceeds a value $g_{\text{c}}(L)$. 

\begin{figure}
 \begin{center}
  (a)\includegraphics[width=9cm,clip]{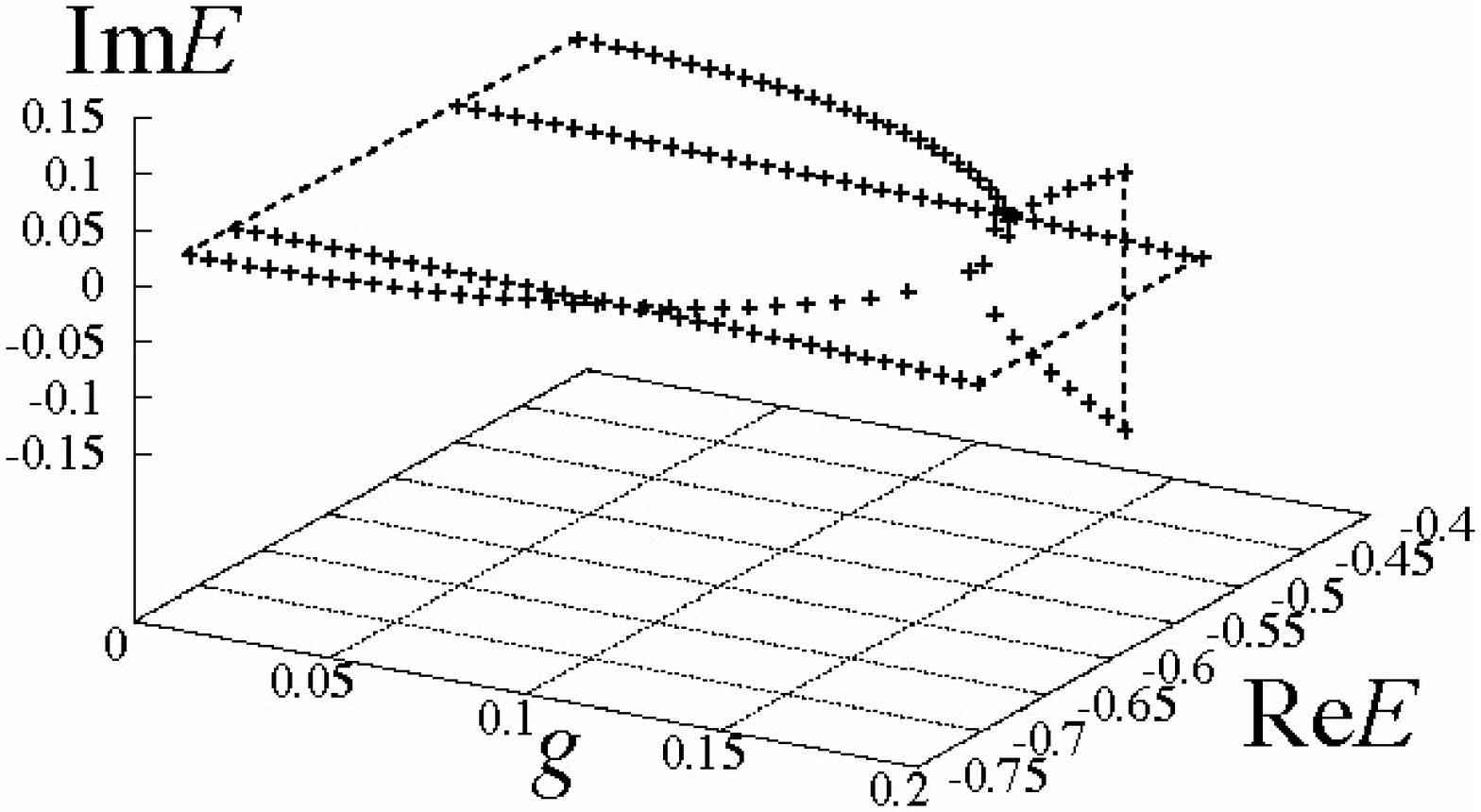} \\
  
 (b)\includegraphics[width=7cm,clip]{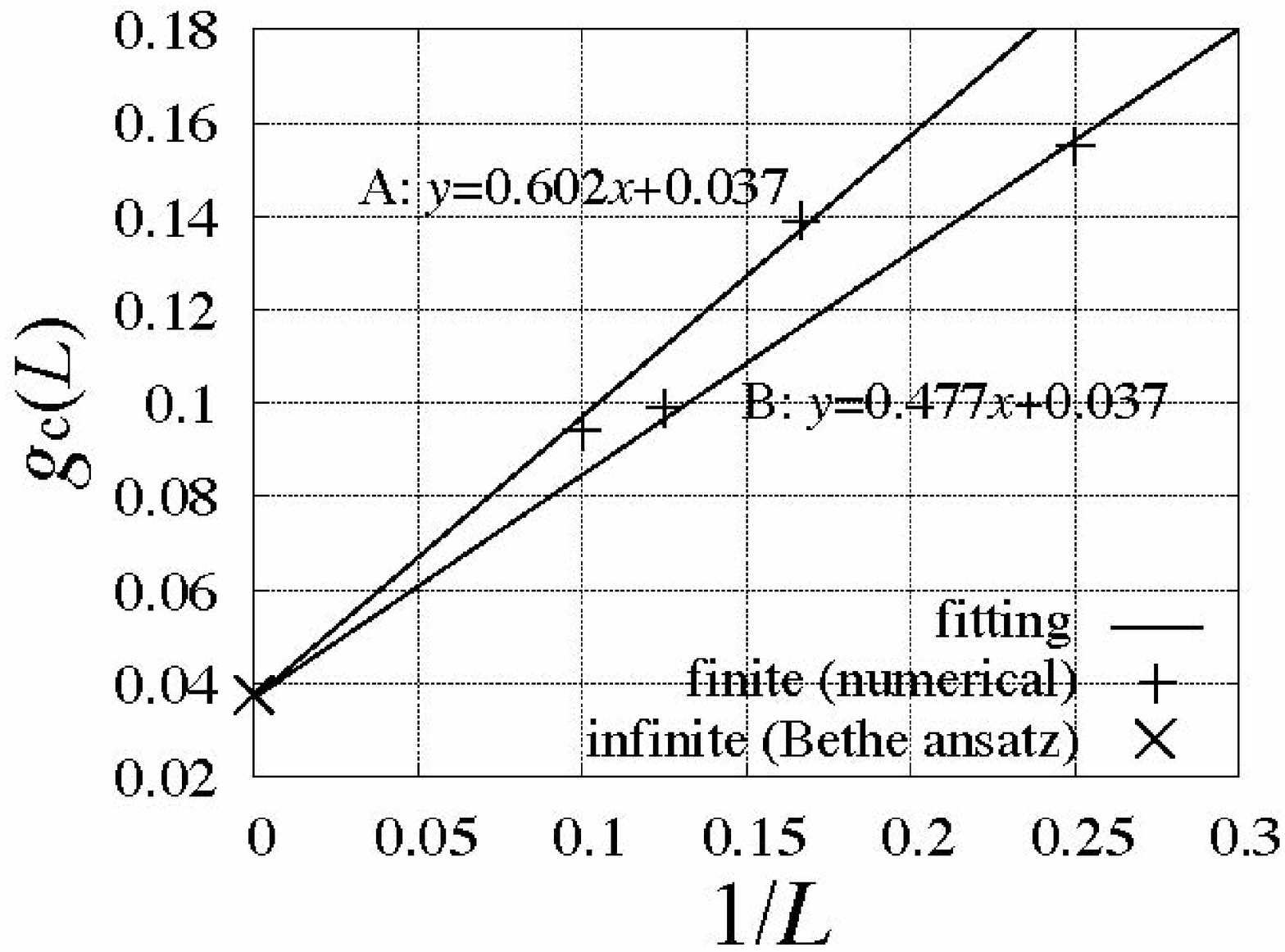} \\
  \caption[]{(a) The spectrum flow of the eigenvalues per site around the ground state for $L=4$ with $U/t=2$ as we increase the non-Hermiticity $g$ which eliminates the charge gap.
(b) The $1/L$ plot of $g_{\text{c}}(L)$.}
  \label{Hub_num_charge}
 \end{center}
 \end{figure}

Figure~\ref{Hub_num_charge}~(a) shows the spectrum flow of the eigenvalues per site for $L=4$ around the ground state for $U=2t$. 
The eigenvalues of the ground state and the third excited state move toward each other on the real axis and spread into the complex plane as soon as the two eigenvalues collide, which gives the non-Hermitian ``critical" point $g_{\text{c}}(L)$. The eigenvalues of the first and the second excited states scarcely move. The movement of the ground-state energy is presumably a finite-size effect; the ground-state energy does not change for $g<g_{\text{c}}$ for the infinite system as shown in Fig.~\ref{gap_g_diagram}. 

We numerically estimated $g_{\text{c}}(L)$ for $L=4, 6, 8$ and $10$ and extrapolated them to $g_{\text{c}}(\infty)$ by considering the finite-size correction as follows:
\begin{equation}
g_{\text{c}}(L)=g_{\text{c}}(\infty)+a/L+\text{O}(1/L^2)\label{scaling}.
\end{equation}
Figure~\ref{Hub_num_charge}~(b) shows the $1/L$ plot of $g_{\text{c}}(L)$; this implies that we have to consider different finite-size corrections in the case $L=4n$ (for $L=4$ and $8$) and in the case $L=4n+2$ (for $L=6$ and $10$).
In Fig.~\ref{Hub_num_charge}~(b), the line~A is the linear fitting of $g_{\text{c}}(L)$ for $L=6$ and $10$ and the line~B is that for $L=4$ and $8$; both lines are determined by the least-squares method under the condition that they have the same intercept $g_{\text{c}}(\infty)$.
The final estimate of $g_{\text{c}}(\infty)$ is $0.037$, while the Bethe-ansatz method yields $g_{\text{c}}=1/\xi_{\text{charge}}\cong 0.038$.
Our estimate is consistent with the exact value. It is quite remarkable to obtain such a good estimate from data for such small $L$.

\begin{figure}
 \begin{center}
  (a)\includegraphics[width=9cm,clip]{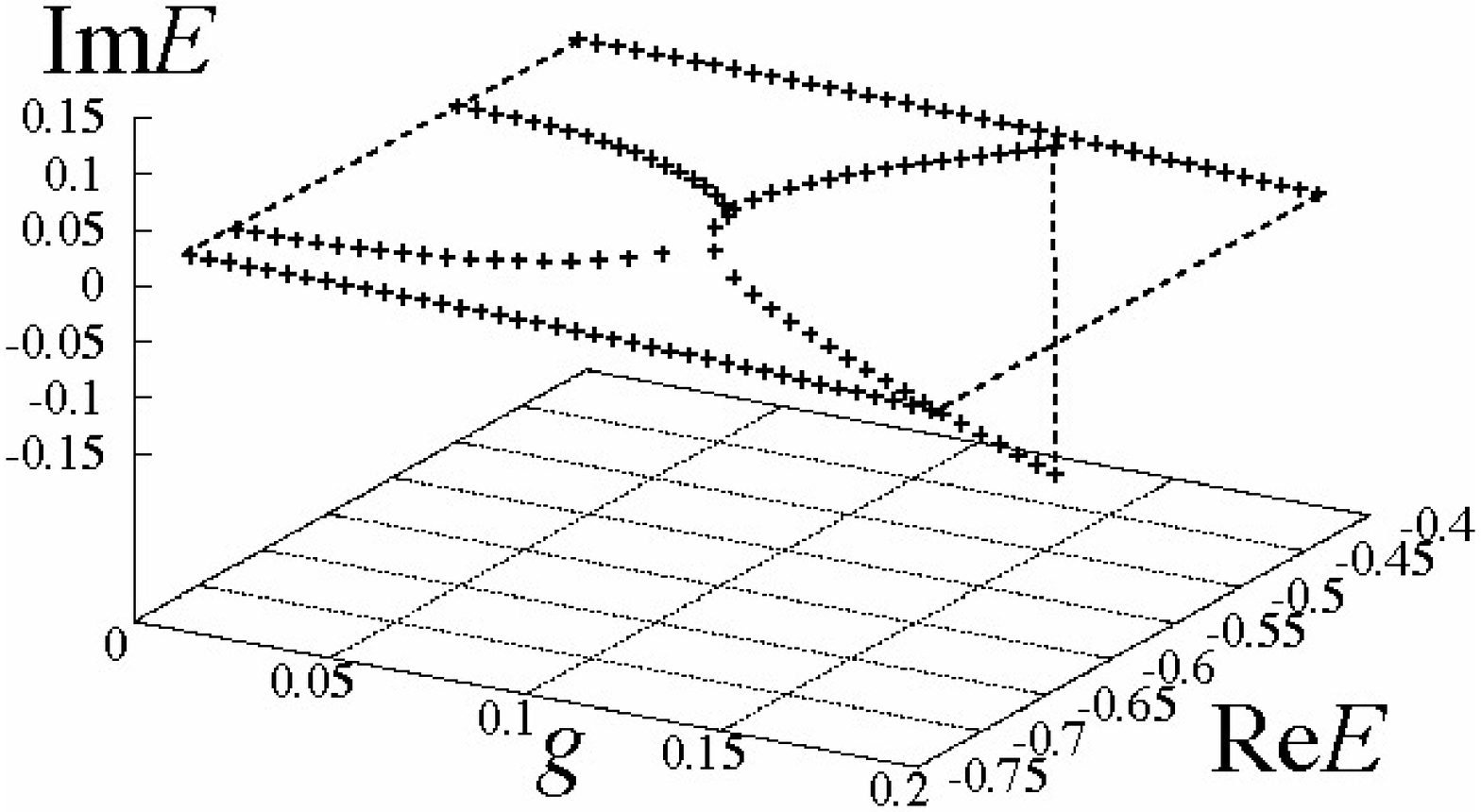} \\
  
 (b)\includegraphics[width=7cm,clip]{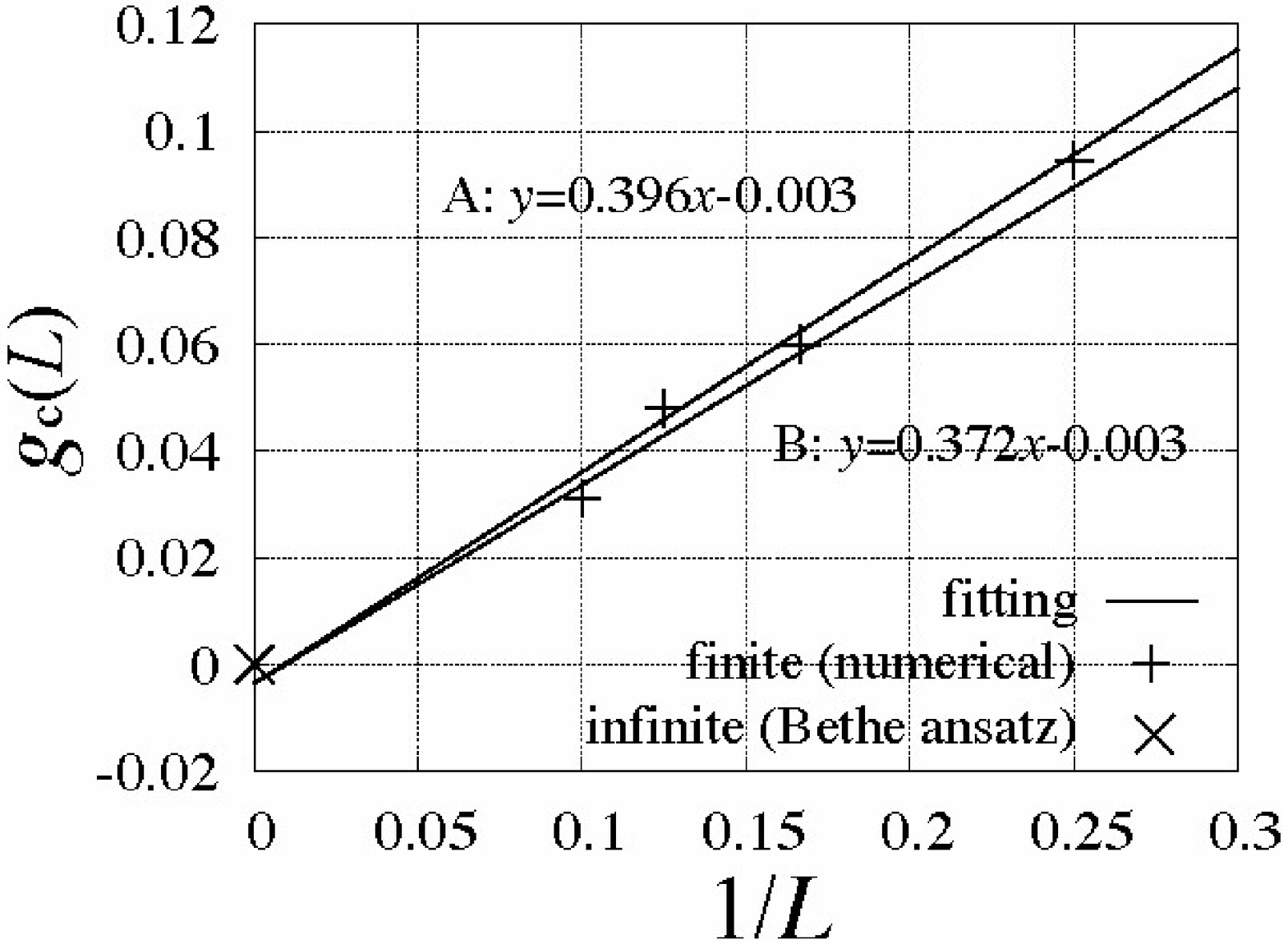} \\
  \caption[]{(a) The spectrum flow of the eigenvalues per site around the ground state for $L=4$ with $U/t=2$ as we increase the non-Hermiticity $g$, which eliminates the spin gap.
(b) The $1/L$ plot of $g_{\text{c}}(L)$.}
  \label{Hub_num_spin}
 \end{center}
 \end{figure}

Figure~\ref{Hub_num_spin}~(a) shows the spectrum flow for $L=4$ around the ground state when we use the non-Hermitian generalization of the form~(\ref{spinon1}), which eliminates the spin gap.
 The eigenvalues of the first and the second excited states move toward each other, while the eigenvalues of the ground state and the third excited state scarcely move. 
We presume that the energy gap between the ground state and the third excited state is caused by the charge excitation while the energy gap between the first and the second excited states is caused by the spinon excitation. We expect that the ground state and the first excited state are eventually degenerate in the thermodynamic limit. Hence we regard the collision of the first and second excited states as the ground-state transition. This behavior implies the charge-spin separation of one-dimensional quantum systems in the low-energy region~\cite{Frahm}. 
Figure~\ref{Hub_num_spin}~(b) shows the $1/L$ plot of $g_{\text{c}}(L)$. We obtain the extrapolated estimate $g_{\text{c}}(\infty)$ by the same least-squares method as we used above; the line~A shows the linear fitting of $g_{\text{c}}(L)$ for $L=4$ and $8$ and the line~B is that for $L=6$ and $10$, which yields the extrapolated estimate $g_{\text{c}}(\infty)=-0.003$. Our estimate is also consistent with the exact value $g_{\text{c}}=1/\xi_{\text{spinon}}=0$.

\subsection{The non-Hermitian $XXZ$ chain} 

\begin{figure}
\begin{center}
(a)\includegraphics[width=7cm,clip]{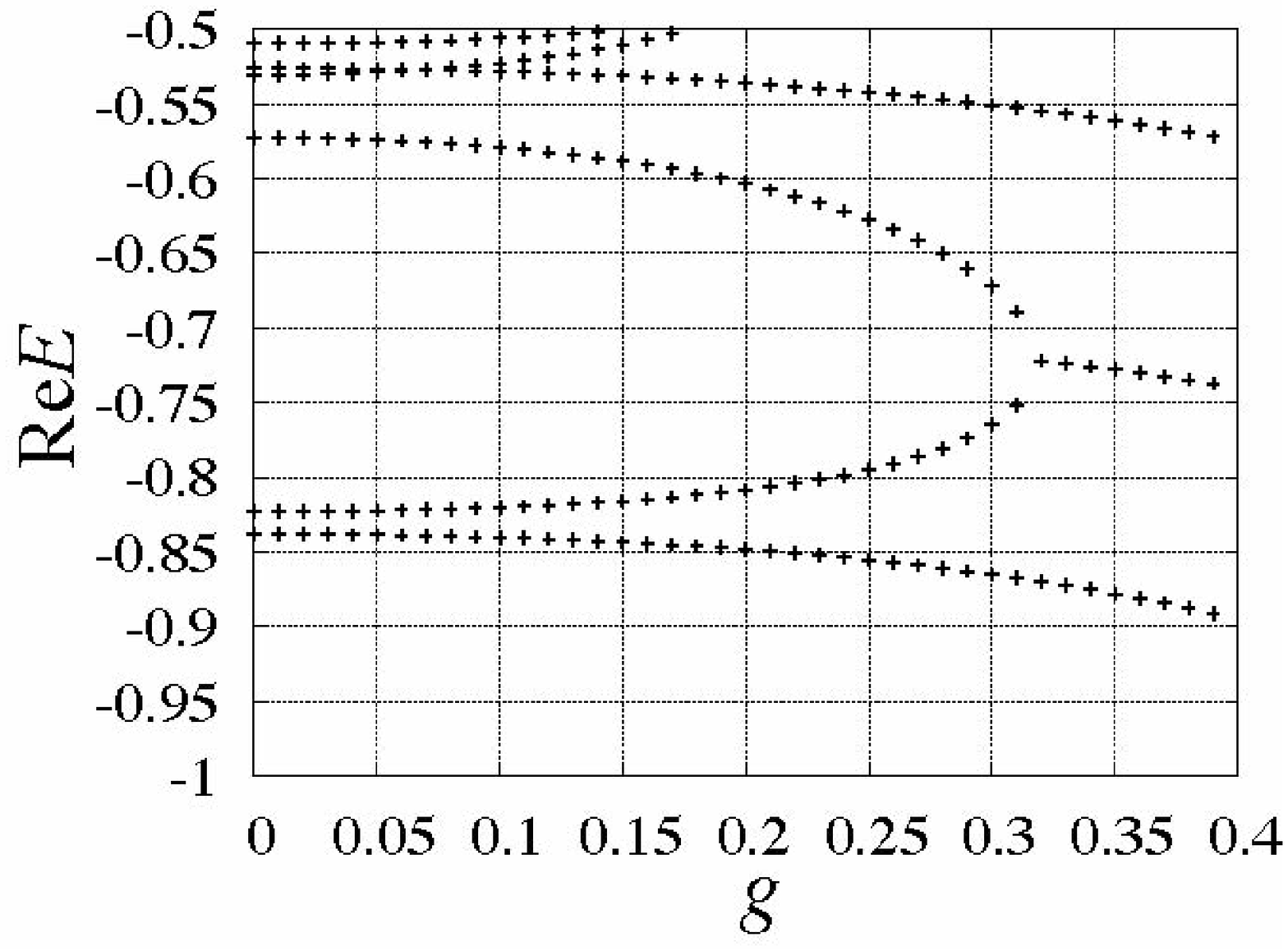}\\

(b)\includegraphics[width=8cm,clip]{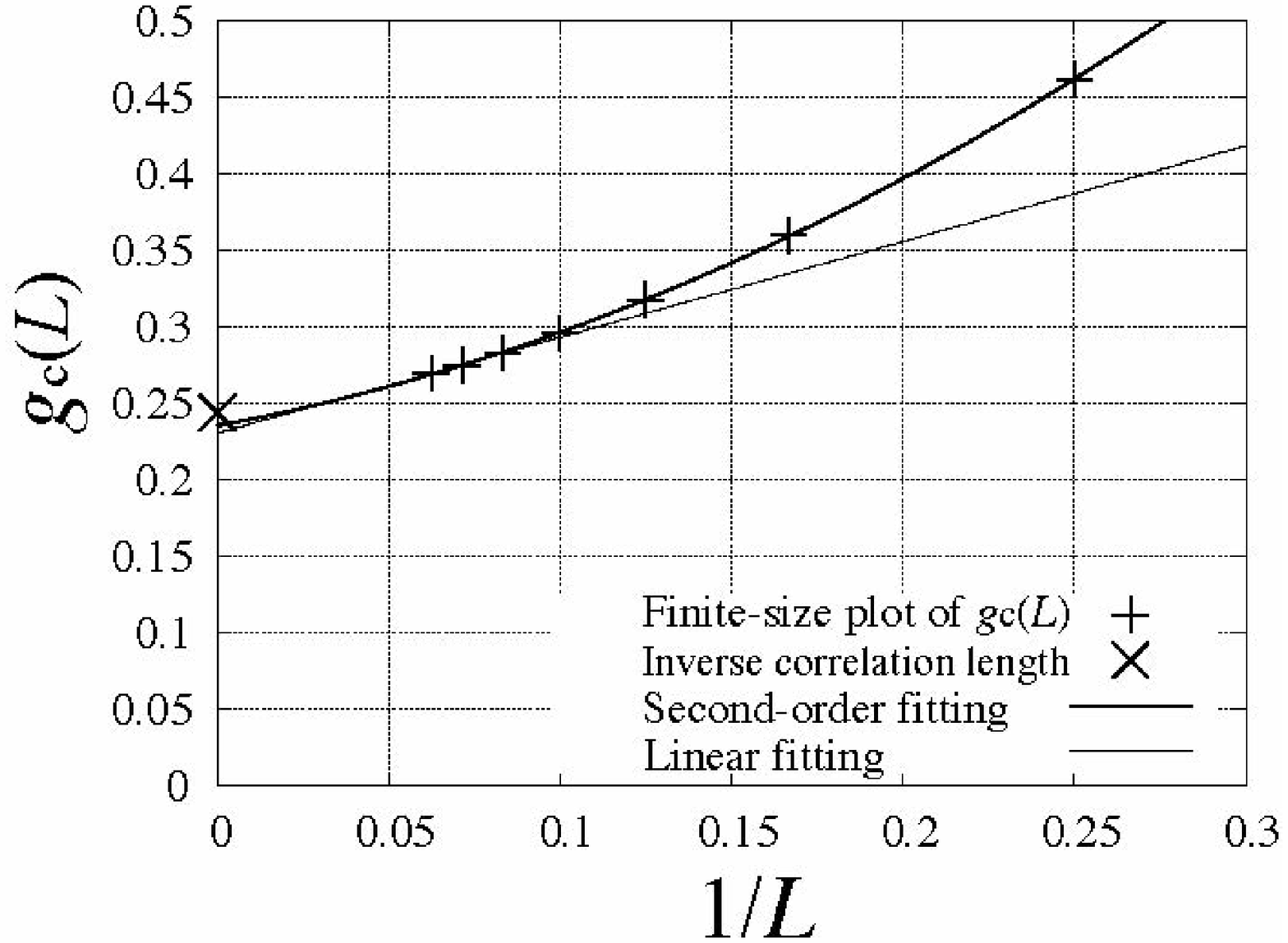} 
  \caption[]{(a) The spectrum flow of the eigenvalues per site of the $XXZ$ model for $L=8$ with $\Delta=3$. (b) The finite-size plot of $g_{\text{c}}(L)$~\cite{Nakamura}.}
\label{XXZ_spectrum_flow}
 \end{center}
 \end{figure}

Next, we numerically calculate the non-Hermitian ``critical" point $g_{\text{c}}(L)$ of the $XXZ$ chain of size $L$. We obtain the extrapolated estimate $g_{\text{c}}(\infty)$ of finite-size data $g_{\text{c}}(L)$ and show that the estimate $g_{\text{c}}(\infty)$ is consistent with the inverse correlation length of the Hermitian system as was for the Hubbard model. Figure~\ref{XXZ_spectrum_flow}~(a) shows the spectrum flow per site around the ground state of the $XXZ$ chain of $L=8$ as we increase the non-Hermiticity $g$ for $\Delta=3$.
The pair of the first and the second excited states undergoes the real-complex transition, which is the same as in Fig.~\ref{Hub_num_spin}~(a). Note that the ground state and the first excited state of Hermitian finite systems are eventually degenerate in the thermodynamic limit. We therefore expect that the real-complex transition point of the first and second excited states converges to the non-Hermitian critical point $g_{\text{c}}$ in the limit $L\to\infty$. We thereby use the real-complex transition point in order to define the non-Hermitian ``critical" point $g_{\text{c}}(L)$.

We then extrapolate the finite-size data.
Figure~\ref{XXZ_spectrum_flow}~(b) shows the extrapolation of $g_{\text{c}}(L)$ for the $XXZ$ chain with $\Delta=3$.
 The extrapolated estimate $g_{\text{c}}(\infty)$ by linear fitting in the form
 \begin{equation}
g_{\text{c}}(L)=g_{\text{c}}(\infty)+a/L+\text{O}(1/L^2)
\end{equation}
 for $L=12, 14$ and $16$, is 0.231. 
 In order to take the finite-size data $g_{\text{c}}(L)$ for small $L$ into consideration, we also calculate the extrapolated estimate $g_{\text{c}}(\infty)$ by a second-order fitting in the form
\begin{equation}
g_{\text{c}}(L)=g_{\text{c}}(\infty)+a/L+b/L^2+\text{O}(1/L^3)
\end{equation}
 for $L=4,6, \dots, 14, 16$ to obtain $g_{\text{c}}(\infty)=0.235$. 
 Both estimates are consistent with the inverse correlation length $g_{\text{c}}=1/\xi\cong 0.244$ calculated analytically.

\subsection{The NNN Heisenberg chain} 

We consider the non-Hermitian generalization of the $S=1/2$ antiferromagnetic Heisenberg chain with nearest- and next-nearest-neighbor interactions; we hereafter call this model NNN Heisenberg chain.
The Hermitian Hamiltonian of this model is
\begin{equation}
\mathcal{H}_{\text{NNN}}=J\sum_{l=1}^{L}[\mathbf{S}_{l}\cdot\mathbf{S}_{l+1}+\alpha\mathbf{S}_{l}\cdot\mathbf{S}_{l+2}],
\end{equation}
where $J>0$ and $\alpha\geq 0$. We require the periodic boundary conditions. At the point $\alpha=0$, the model is the standard Heisenberg chain. The ground state is a spin fluid state and the energy gap is zero. At the point $\alpha=1/2$, the model is the Majumdar-Ghosh model~\cite{MG1, MG2} and the energy gap is finite. Okamoto and Nomura~\cite{Okamoto} numerically showed that a massive-massless transition occurs at $\alpha_{\mathrm{c}}\cong 0.2411$. 

\begin{figure}
\begin{center}
  (a)\includegraphics[width=8cm,clip]{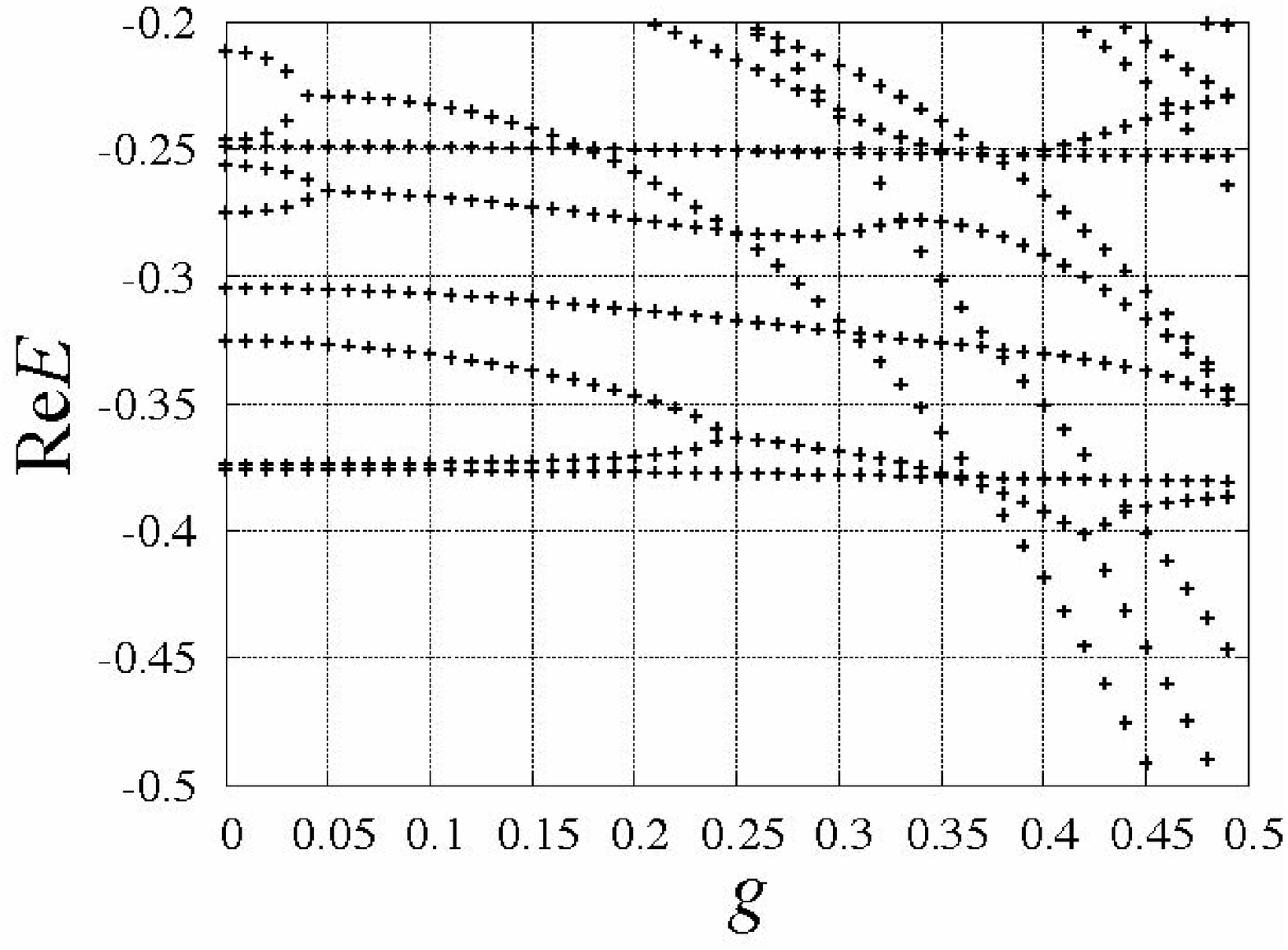}\\ 
  
 (b)\includegraphics[width=8cm,clip]{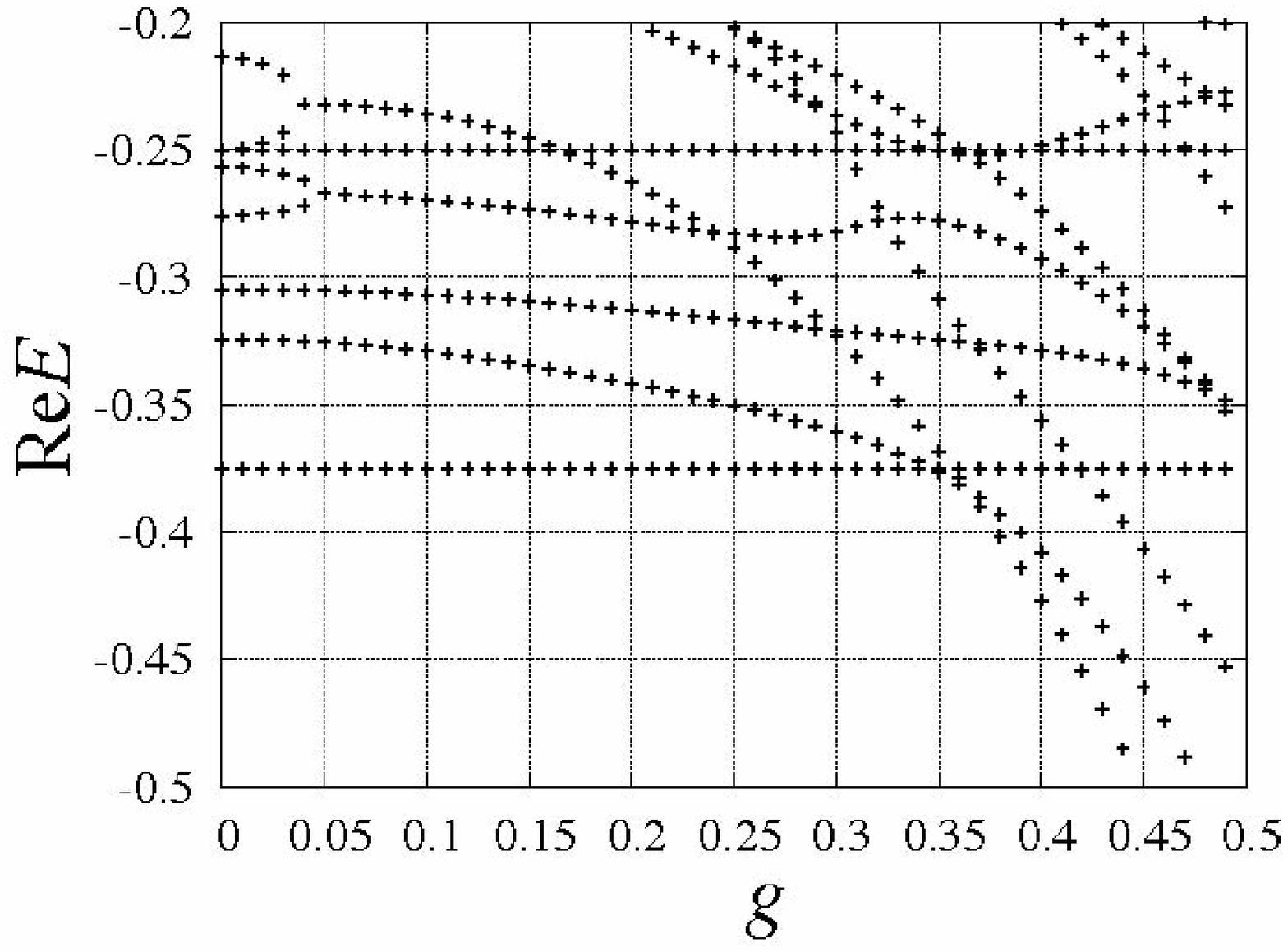}\\
 
  (c)\includegraphics[width=8cm,clip]{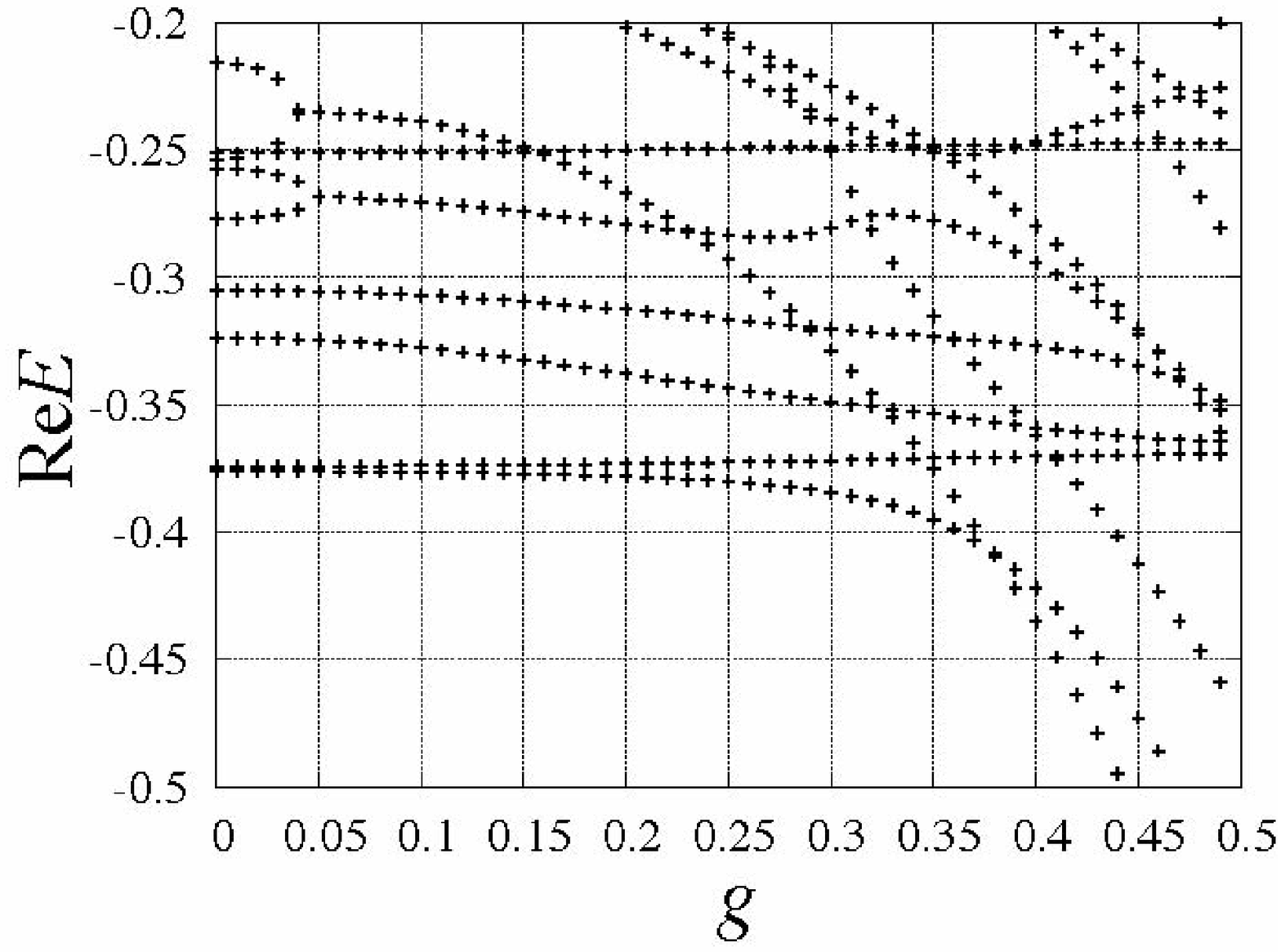} 
  \caption[]{The spectrum flow of the real part of the eigenvalues per site around the ground state for $L=8$ with (a) $\alpha=0.49$, (b) $\alpha=0.5$ and (c) $\alpha=0.51$.}
  \label{zigzag_flow}
 \end{center}
 \end{figure}

We here analyze the non-Hermitian Hamiltonian of the NNN Heisenberg chain
\begin{equation}
\mathcal{H}_{\text{NNN}}=J\sum_{l=1}^{L}[\frac{1}{2}(e^{2g}S_{l+1}^{+}S_{l}^{-}+e^{-2g}S_{l}^{+}S_{l+1}^{-})+S_{l}^{z}S_{l+1}^{z}]+\alpha J\sum_{l=1}^{L}[\frac{1}{2}(e^{4g}S_{l+2}^{+}S_{l}^{-}+e^{-4g}S_{l}^{+}S_{l+2}^{-})+S_{l}^{z}S_{l+2}^{z}].
\label{NNN_Hamiltonian}
\end{equation} 
We numerically estimated the non-Hermitian ``critical" point $g_{\text{c}}(L)$ of finite systems in the subspace $S_{\text{tot}}^{z}=0$ where an eigenvalue which corresponds to the ground state in the thermodynamic limit becomes complex. Figure~\ref{zigzag_flow} shows the spectrum flow of the eigenvalues per site around the ground state for $L=8$ with $\alpha=0.49$, $\alpha=0.5$~(the Majumdar-Ghosh point) and $\alpha=0.51$. 
In Fig.~\ref{zigzag_flow}~(a) for $\alpha=0.49$, as we increase the non-Hermiticity $g$, the energy gap between the first excited state, which corresponds to one of the degenerate ground states in $L\to\infty$, and the second excited state, which corresponds to the first excited state in $L\to\infty$, vanishes at $g=g_{\text{c1}}(L)\cong 0.24$. These two eigenvalues become complex in the region $g_{\text{c1}}(L)<g<g_{\text{c2}}(L)$ before they become real again at $g=g_{\text{c2}}(L)\cong 0.42$. We define the non-Hermitian ``critical" point of a finite system for $\alpha<0.5$ as the point $g=g_{\text{c1}}(L)$ where the first excited-state energy first becomes complex. 

In Fig.~\ref{zigzag_flow}~(b) for $\alpha=0.5$, the two-fold degenerate ground states exist for any $g$ and the energy gap between the ground state and the first excited state vanishes at $g=g_{\text{c}}(L)\cong 0.35$. However, these eigenvalues do not become complex for $g>g_{\text{c}}(L)$. This is presumably because  $g_{\text{c1}}(L)=g_{\text{c2}}(L)=g_{\text{c}}(L)$. We still regard $g_{\text{c1}}(L)$ as the non-Hermitian ``critical" point $g_{\text{c}}(L)$ for $\alpha=0.5$. 

In Fig.~\ref{zigzag_flow}~(c) for $\alpha=0.51$, the ground state and the first excited state never become complex for any $g$. The ground state for $\alpha>0.5$ is an incommensurate state of the spiral phase~\cite{Chitra}. Our non-Hermitian generalization of the form~(\ref{NNN_Hamiltonian}) is presumably not appropriate for detecting the incommensurate state in the region $\alpha>0.5$; we may need other types of non-Hermitian generalization. 

\begin{figure}
 \begin{center}
 (a)\includegraphics[width=8cm,clip]{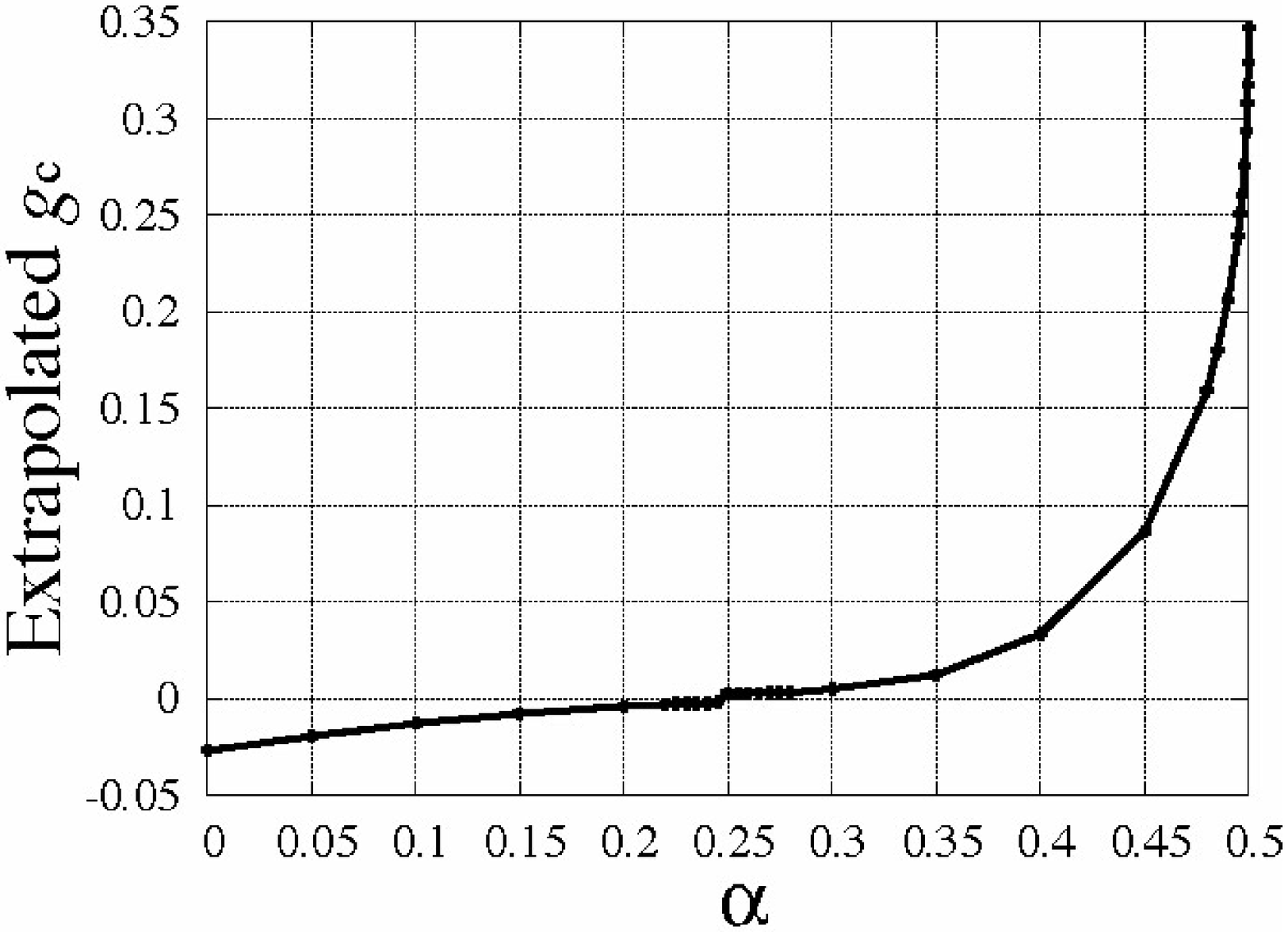} \\
 
 (b)\includegraphics[width=8cm,clip]{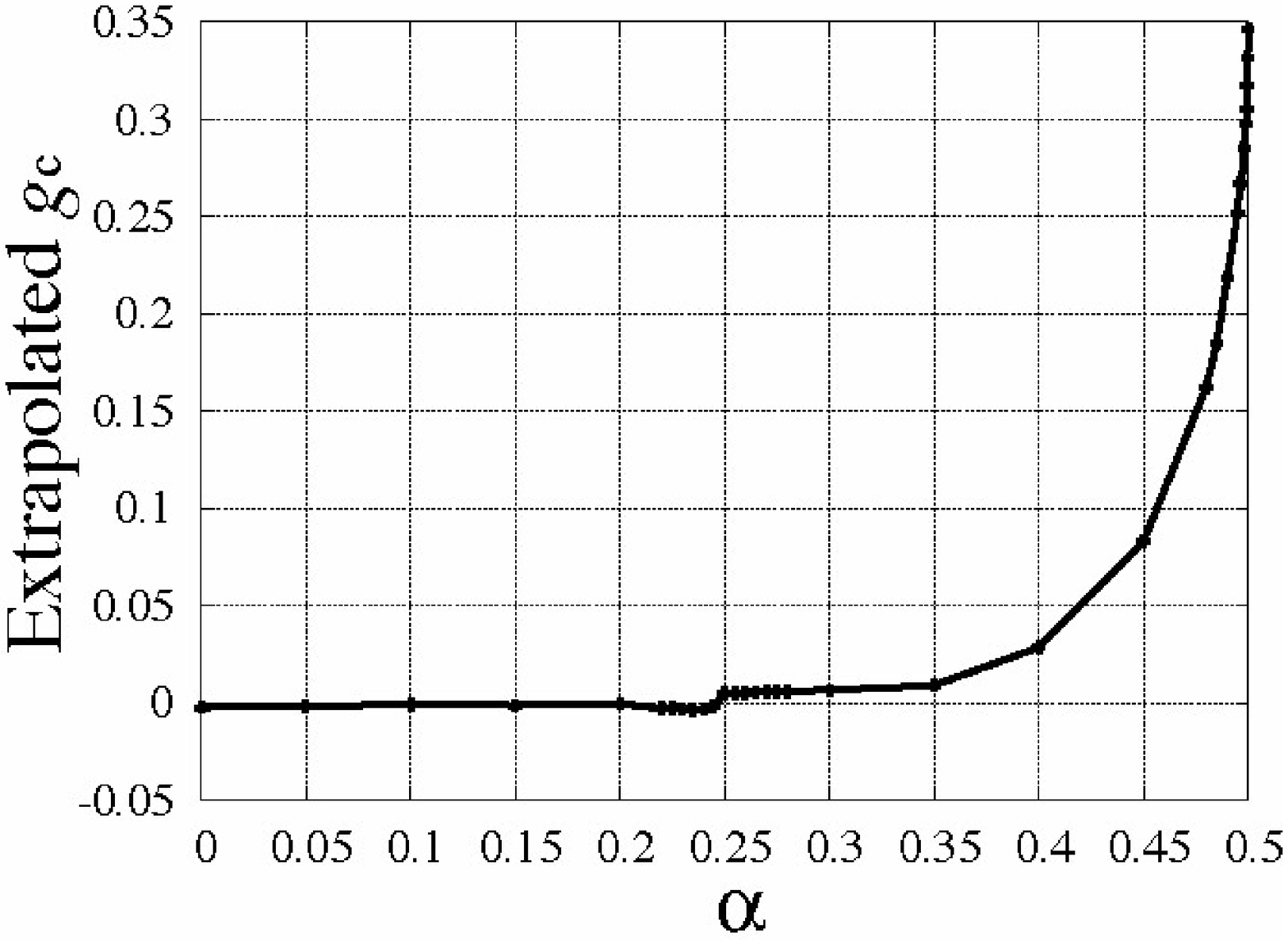} 
  \caption[]{The extrapolated estimates $g_{\text{c}}(\infty)$ as a function of $\alpha$ by (a) a linear fitting and by (b) a second-order fitting.}
  \label{zigzag_gc}
 \end{center}
 \end{figure}

We hereafter restrict ourselves to the region $0\leq\alpha\leq 0.5$. We extrapolate the finite-size data $g_{\text{c}}(L)$ for $L=4,8,12$ and $16$ by the linear fitting in the form
\begin{equation}
g_{\text{c}}(L)=g_{\text{c}}(\infty)+a/L+\text{O}(1/L^2)
\end{equation}
and by the second-order fitting in the form
\begin{equation}
g_{\text{c}}(L)=g_{\text{c}}(\infty)+a/L+b/L^2+\text{O}(1/L^3).
\end{equation}
Figure~\ref{zigzag_gc} shows the extrapolated estimates $g_{\text{c}}(\infty)$ in the region $0\leq\alpha\leq 0.5$.
The second-order estimate $g_{\text{c}}(\infty)$ around $\alpha\cong 0$ is almost zero and is consistent with $\xi^{-1}=0$ in the limit $L\to\infty$.
The extrapolated estimates $g_{\text{c}}(\infty)$ at $\alpha=0.5$ are around 0.35 for both linear and second-order fitting, which is consistent with the inverse correlation length $\xi^{-1}=\ln 2/2~(\cong 0.347)$; see \S.~2.4. 
The estimate $g_{\text{c}}(\infty)$ is almost zero in the region $0\leq\alpha\lesssim 0.25$ and is finite  in the region $0.25\lesssim\alpha\leq 0.5$. This is consistent with the massive-massless transition~\cite{Okamoto} at $\alpha\cong 0.2411$ if we admit that $g_{\text{c}}(\infty)$ is equal to the inverse correlation length. 
It is remarkable that $g_{\text{c}}(\infty)$ is at least approximately equal to the inverse correlation length all through the region $0\leq\alpha\leq 0.5$ though we do not know the exact dispersion relation of the elementary excitation of this model. 

\section{Summary and discussions}
To summarize, we considered the non-Hermitian generalization of strongly correlated quantum systems, where we make the hopping energy of the original Hermitian Hamiltonian asymmetric. We rigorously showed that the non-Hermitian critical point where the energy gap vanishes is equal to the inverse correlation length of the Hermitian systems for the ferromagnetic isotropic \textit{XY} chain in the magnetic field, the half-filled Hubbard model, the $S=1/2$ antiferromagnetic \textit{XXZ} chain in the Ising-like region and the Majumdar-Ghosh model. We argued that the non-Hermitian generalization replaces the real momentum $k$ with the complex one $k+ig$ or $k+2ig$ in the dispersion relation of the elementary excitation. By the non-Hermitian generalization, we effectively seek the zero of the dispersion relation in the complex momentum space; the imaginary part of the momentum at the zero is related to the inverse correlation length of the Hermitian system.

We also discussed the $S=1/2$ antiferromagnetic Heisenberg chain with the nearest- and the next-nearest-neighbor interactions, which is unsolved analytically.
We presented numerical evidence that the extrapolated estimates of the non-Hermitian ``critical" point for finite systems is consistent with the inverse correlation length.

The non-Hermitian generalization of the form~(\ref{non-Herm-generalization}) is not always appropriate for having the inverse correlation length. The model~(\ref{NNN_Hamiltonian}) with $\alpha>0.5$ is an example. The $S=1/2$ ferromagnetic transverse Ising chain
\begin{equation}
\mathcal{H}=-J\sum_{l=1}^{L}S_{l}^{x}S_{l+1}^{x}-h\sum_{l=1}^{L}S_{l}^{z}
\label{Hamiltonian_Trans_Ising}
\end{equation}
is another example.
The elementary excitation of eq.~(\ref{Hamiltonian_Trans_Ising}) is given by
\begin{equation}
\eta_{k}=\cos\theta_{k}c_{k}-\sin\theta_{k}c_{-k}^{\dag},\hspace{9pt}\eta_{-k}=\sin\theta_{k}c_{k}^{\dag}+\cos\theta_{k}c_{-k},
\end{equation}
where 
\begin{equation}
\theta_{k}=-\frac{1}{2}\arctan\left[\frac{J\sin k}{J\cos k+2h}\right].
\end{equation}
It has the dispersion relation
\begin{equation}
\epsilon(k)=\sqrt{(J\cos k/2+h)^2+(J\sin k/2)^2}.
\end{equation}
By replacing $k$ with $k+ig$ in the dispersion relation $\epsilon(k)$, we obtain the non-Hermitian Hamiltonian in the form~(\ref{dispersion_non-Hermitian}).
This is transformed back to the spin Hamiltonian of the form
\begin{align}
\mathcal{H}(g)=&\sum_{n=1}^{\infty}\sum_{l=1}^{L}(-2)^{n}S_{l+1}^{z}\dots S_{l+n-1}^{z}\left[(p_{n}-r_{n})S_{l+n}^{x}S_{l}^{x}+(p_{n}+r_{n})S_{l+n}^{y}S_{l}^{y}+iq_{n}(S_{l+n}^{x}S_{l}^{y}-S_{l+n}^{y}S_{l}^{x})\right]\notag\\
&-p_{0}\sum_{l=1}^{L}S_{l}^{z},
\end{align}
where coefficients $p_{n}, q_{n}$ and $r_{n}$ are given by the following integrals:
\begin{align}
p_{n}&=\frac{1}{2\pi}\int_{-\pi}^{\pi}\left(\frac{J\cos k}{2}+h\right)\frac{\epsilon(k+ig)+\epsilon(k-ig)}{2\epsilon(k)}\cos(nk)dk,  \notag\\
q_{n}&=\frac{i}{2\pi}\int_{-\pi}^{\pi}\frac{\epsilon(k+ig)-\epsilon(k-ig)}{2}\sin(nk)dk,  \notag\\
r_{n}&=-\frac{1}{2\pi}\int_{-\pi}^{\pi}\frac{J\sin k}{2}\frac{\epsilon(k+ig)+\epsilon(k-ig)}{2\epsilon(k)}\sin(nk)dk.
\end{align}
This non-Hermitian Hamiltonian is very complicated; interactions among spins beyond the nearest neighbor sites emerge as soon as $g$ is finite.  It is because its elementary excitation is obtained by the Bogoliubov transformation; the creation and annihilation operators at two different wave numbers $k$ and $-k$ are mixed. Conversely, the non-Hermitian generalization of the simple form~(\ref{non-Herm-generalization}) does not produce the dispersion relation $\epsilon(k+ig)$ in this model. We may need another principle of non-Hermitian generalization.

\begin{acknowledgments}
The authors would like to thank Akinori Nishino and Masahiro Shiroishi for helpful discussions. The work is supported by Grant-in-Aid for Scientific Research~(No.~17340115) from the Ministry of Education, Culture, Sports, Science and Technology as well as by Core Research for Evolutional Science and Technology of Japan Science and Technology Agency.
\end{acknowledgments}

\appendix
\section{Equality of $g_{\text{c}}$ and $1/\xi$ for the Hubbard model}
We show for the Hubbard model that the non-Hermitian critical point $g_\text{c}$ in eq.~(\ref{gc_Hubbard}) and the inverse correlation length $1/\xi$ in eq.~(\ref{xi_Hubbard}) are actually equal.
The non-Hermitian critical point $g_{\text{c}}$ is given by
\begin{align}
g_{\text{c}}&=\lim_{\Lambda\to\infty}\left[\text{arcsinh}(U/4t)+2i\int_{-\Lambda}^{\Lambda}\arctan\frac{\lambda+iU/4t}{U/4t}\sigma(\lambda)d\lambda\right]
\notag\\
&=\lim_{\Lambda\to\infty}\left[\text{arcsinh}(U/4t)+\frac{i}{\pi}\int_{-\Lambda}^{\Lambda}d\lambda\arctan\frac{\lambda+iU/4t}{U/4t}\int_{0}^{\infty}\frac{\cos(\omega\lambda)J_{0}(\omega)}{\cosh((U/4t)\omega)}d\omega\right]. \label{gc_Hubbard_appendix}
\end{align}
Using the variable transformation
\begin{equation}
\theta=\arctan(\lambda/(U/4t)+i)
\end{equation}
with 
\begin{equation}
\tan\theta_1=-\frac{\Lambda}{U/4t}+i,\hspace{9pt}\tan\theta_2=\frac{\Lambda}{U/4t}+i,
\end{equation}
we have
\begin{align}
g_{\text{c}}=&\lim_{\Lambda\to\infty}\left[\text{arcsinh}(U/4t)+\frac{i}{\pi}\int_{\theta_{1}}^{\theta_{2}}\frac{(U/4t)\theta}{\cos^{2}\theta}\left[\int_{0}^{\infty}\frac{\cos((U/4t)\omega\tan\theta-i(U/4t)\omega)J_{0}(\omega)d\omega}{\cosh((U/4t)\omega)} \right]d\theta\right] \notag\\
=&\lim_{\Lambda\to\infty}\left[\text{arcsinh}(U/4t)+\frac{i}{\pi}\int_{0}^{\infty}\frac{(U/4t)J_{0}(\omega)d\omega}{\cosh((U/4t)\omega)}\right.\notag\\
&\left.\times\left\{\underbrace{\left[\frac{\sin((U/4t)\omega\tan\theta-i(U/4t)\omega)}{(U/4t)\omega}\theta \right]_{\theta_{1}}^{\theta_{2}}}_{I_{1}} \right.\left.
-\underbrace{\int_{\theta_{1}}^{\theta_{2}}\frac{\sin((U/4t)\omega\tan\theta-i(U/4t)\omega)}{(U/4t)\omega}d\theta}_{I_{2}}   \right \}\right].\label{gc_equality}
\end{align}
We rewrite the term $I_1$ in the form
\begin{align}
I_1=&\frac{1}{(U/4t)\omega}\left[\theta_{2}\sin((U/4t)\omega\tan\theta_{2}-i(U/4t)\omega)-\theta_{1}\sin((U/4t)\omega\tan\theta_{1}-i(U/4t)\omega)\right]\notag\\
=&\frac{1}{(U/4t)\omega}(\theta_{1}+\theta_{2})\sin(\omega\Lambda),\label{eq4_Bethe11}
\end{align}
where the coefficients $\theta_{1}$ and $\theta_{2}$ for $\Lambda\gg 1$ take the form
\begin{equation}
\theta_{1}=-\frac{\pi}{2}-\delta_{1}, \hspace{9pt}\theta_{2}=\frac{\pi}{2}-\delta_{2}
\end{equation}
with $|\delta_{1}|, |\delta_{2}|\ll 1$. Because of
\begin{equation}
\tan\theta_{1}=\frac{1}{\tan\delta_{1}}\simeq\frac{1}{\delta_{1}},\hspace{9pt}
\tan\theta_{2}=\frac{1}{\tan\delta_{2}}\simeq\frac{1}{\delta_{2}},
\end{equation}
we have 
\begin{align}
I_1
&=\frac{1}{(U/4t)\omega}(-\delta_{1}-\delta_{2})\sin(\omega\Lambda)
\notag\\
&=\frac{1}{(U/4t)\omega}\left(-\frac{U/4t}{i(U/4t)-\Lambda}-\frac{U/4t}{i(U/4t)+\Lambda}\right)\sin(\omega\Lambda)\notag\\
&=\frac{2i(U/4t)}{\omega((U/4t)^2+\Lambda^2)}\sin(\omega\Lambda)\notag\\
&\xrightarrow{\Lambda\to\infty} 0.
\end{align}
Next we calculate the integral $I_2$ in eq.~(\ref{gc_equality}).
By using the variable transformation $x=\tan\theta-i$, we have 
\begin{align}
I_2&=\lim_{\Lambda \to \infty}\int_{-\Lambda/(U/4t)}^{\Lambda/(U/4t)}\frac{\sin((U/4t)\omega x)}{(U/4t)\omega}\frac{dx}{1+(x+i)^2}\notag\\
&=\int_{-\infty}^{\infty}\frac{\sin((U/4t)\omega x)}{(U/4t)\omega(x^2+2ix)}dx
=\frac{\pi}{2i(U/4t)\omega}(1-e^{-2\omega(U/4t)}).
\end{align}
We thus arrive at
\begin{align}
g_{\text{c}}=&\text{arcsinh}(U/4t)-\frac{i}{\pi}\int_{0}^{\infty}\frac{(U/4t)J_{0}(\omega)}{\cosh((U/4t)\omega)}\frac{\pi}{2i(U/4t)\omega}(1-e^{-2\omega(U/4t)})d\omega \notag\\
=&\text{arcsinh}(U/4t)-2\int_{0}^{\infty}\frac{J_{0}(\omega)\sinh((U/4t)\omega)}{\omega(1+e^{2(U/4t)\omega})}d\omega. \label{eq4_critical1}
\end{align}
We thus confirmed that the expression~(\ref{eq4_critical1}), or eq.~(\ref{gc_Hubbard}) is actually equal to the inverse correlation length of the charge excitation~(\ref{xi_Hubbard}).

\section{A zero of the dispersion relation of the elementary excitation and the correlation length}

In \S~2.1, we showed eq.~(\ref{dispersion_XY}) for the \textit{XY} chain.
In this section, we argue similar relations for the half-filled Hubbard model and the $S=1/2$ antiferromagnetic \textit{XXZ} chain. That is, the imaginary part of the momentum at a zero of the dispersion relation of the elementary excitation in the \textit{complex} momentum space is related to the correlation length.

\subsection{The half-filled Hubbard chain} 
 We consider the Hermitian Hubbard model
 \begin{equation}
\mathcal{H} = -t\sum_{l=1,\sigma=\uparrow,\downarrow}^{L} (c_{l+1,\sigma}^\dag c_{l,\sigma}+c_{l,\sigma}^\dag c_{l+1,\sigma})+ U\sum_{l=1}^{L} c_{l,\uparrow}^\dag c_{l,\uparrow}c_{l,\downarrow}^\dag c_{l,\downarrow}\label{Hubbard_Hermitian}
\end{equation}
in the half-filled case.
The charge excitation at the quasimomentum $k=k_{h}$ has the excitation energy in the form 
\begin{equation}
\mathcal{E}(k_h)=U+4t\cos k_h+8t\int_{0}^{\infty}\frac{\cos(\omega\sin k_h)J_{1}(\omega)}{\omega(1+e^{U\omega/2t})}d\omega
\end{equation}
and has the momentum in the form
\begin{equation}
p(k_h)=k_h+2\int_{0}^{\infty}\frac{\sin(\omega\sin k_h)J_{0}(\omega)}{\omega(1+e^{U\omega/2t})}d\omega.
\end{equation}
By analytic continuation of the dispersion relation of the Hermitian model, we search a zero of $\mathcal{E}(k_h)$ along the axis $\mathop{\textrm{Re}}p(k_h)=\pm\pi$ in the complex momentum space, which gives the Hubbard gap.
Fukui and Kawakami~\cite{Fukui} pointed out that $\mathcal{E}(k_h)$ vanishes at $k_h=\pm\pi+i\text{arcsinh}(U/4t)$.
The momentum $p(k_h)$ at this point is
\begin{equation}
p(k_h)=\pm\pi+i\left[\text{arcsinh}\left(\frac{U}{4t}\right)-2\int_{0}^{\infty}\frac{J_{0}(\omega)\sinh(U/4t)}{\omega(1+e^{\omega U/2t})} d\omega\right].\label{mom_Hubbard}
\end{equation}
We note that the imaginary part of the momentum (\ref{mom_Hubbard}) is equal to the inverse correlation length $1/\xi$ given in eq.~(\ref{xi_Hubbard}). That is, we have $p(k_h)=\pm\pi+i/\xi$ when $\mathcal{E}(k_h)=0$.
This is the same situation as in eq.~(\ref{dispersion_XY}) for the \textit{XY} chain.

\subsection{The $S=1/2$ antiferromagnetic \textit{XXZ} chain} 
We next consider the $S=1/2$ antiferromagnetic \textit{XXZ} chain 
\begin{equation}
\mathcal{H}=J\sum_{l=1}^{L}\left[\frac{1}{2}\left(S_{l}^{-}S_{l+1}^{+}+S_{l}^{+}S_{l+1}^{-}\right)+\Delta S_{l}^{z}S_{l+1}^{z} \right]  \label{XXZ_Hermitian}
\end{equation}
in the region $\Delta>1$. 
The spinon excitation at the quasimomenta $\lambda=\lambda_{s_1}$ and $\lambda=\lambda_{s_2}$ has the excitation energy in the form
\begin{equation}
\mathcal{E}(\lambda_{s_1}, \lambda_{s_2})=\frac{2\pi J\sinh\gamma}{\gamma}\left(\rho(\lambda_{s_1})+\rho(\lambda_{s_2})\right),
\end{equation}
where the distribution function $\rho(\lambda)$ of the rapidity $\lambda$ is given by
\begin{equation}
\rho(\lambda)=\frac{\gamma}{2\pi}\sum_{n=-\infty}^{n=\infty}\frac{e^{in\gamma\lambda}}{2\cosh(n\gamma)}
\end{equation}
and we set $\gamma=\text{arccosh}\Delta$.
The momentum of the spinon excitation takes the form
\begin{equation}
p(\lambda_{s_1}, \lambda_{s_2})=\frac{\gamma(\lambda_{s_1}+\lambda_{s_1})}{2}+\sum_{n=1}^{\infty}\frac{\sin(n\gamma\lambda_{s_1})+\sin(n\gamma\lambda_{s_2})}{n\cosh(n\gamma)}.
\end{equation}
By analytic continuation of the dispersion relation of the Hermitian model, we search a zero of $\mathcal{E}(\lambda_{s_1}, \lambda_{s_2})$ along the axis $\mathop{\textrm{Re}}p(\lambda_{s_1}, \lambda_{s_2})=\pm\pi$ in the complex momentum space, which gives the spinon energy gap.
Albertini \textit{et al.}~\cite{Albertini} pointed out that $\mathcal{E}(\lambda_{s_1}, \lambda_{s_2})$ vanishes at $\lambda_{s_1}=\lambda_{s_2}=\pm\pi/\gamma+i$.
The momentum $p(\lambda_{s_1}, \lambda_{s_2})$ at this point is
\begin{equation}
p(\lambda_{s_1}, \lambda_{s_2})=\pm\pi+i\left[\gamma+2\sum_{n=1}^{\infty}\frac{(-1)^n\tanh(n\gamma)}{n}\right].\label{mom_XXZ}
\end{equation}
We note that the imaginary part of the momentum (\ref{mom_XXZ}) is equal to twice the inverse correlation length $1/\xi$ given in eq.~(\ref{gc_XXZ}). That is, we have $p(\lambda_{s_1}, \lambda_{s_2})=\pm\pi+2i/\xi$ when $\mathcal{E}(\lambda_{s_1}, \lambda_{s_2})=0$.

\newpage 

\end{document}